\documentclass[preprint]{aastex}
\usepackage{lscape}
\usepackage{longtable}
\setlongtables
\shorttitle{Questioning the pulsation paradigm of LPVs}
\shortauthors{P. Berlioz-Arthaud}

\begin{document}

\title{Long Period Variables: questioning the pulsation paradigm}

\author{P. Berlioz-Arthaud}
\affil{Centre de Recherche Astrophysique de Lyon, CNRS/UMR5574, 69561 St-Genis-Laval, France.}
\email{paul.berlioz-arthaud@univ-lyon1.fr}

\begin{abstract}
Long period variables, among them Miras, are thought to be pulsating. Under this approach the whole star inflates and deflates along a period that can vary from 100 to 900 days; that pulsation is assumed to produce shock waves on the outer layers of the star that propagate into the atmosphere and could account for the increase in luminosity and the presence of emission lines in the spectra of these stars. However, this paradigm can seriously be questioned from a theoretical point of view. First, in order to maintain a radial pulsation, the spherical symmetry of the star must be preserved: how can it be reconciled with the large convective cells present in these stars? or when close companions are detected? Secondly, how different radial  and non-radial pulsation modes of a sphere could be all damped except one radial mode? These problems have no solution and significantly weigh on the pulsation paradigm. Acknowledging this inconsistency, we show that a close companion around these stars could account for the star variability. To support this assertion we study the observed light curves, their shapes at different wavelengths and their changes over time.
\end{abstract}


\section{Introduction}

More than 400 years ago, Fabricius observed a star of the Ceti constellation, fading, vanishing and reappearing later with a period which appeared to be of 330 days. 

Soon after however, other variables were discovered, first an eclipsing variable, named Algol, then other Miras: $\chi$ Cyg, R Hya at the beginning of the 18th century. To this day the \emph{``General Catalogue of Variable Stars''} lists more than 8000 Mira type variables. If the mechanism of Algol variability was quickly explained by a mutual eclipse of two stars orbiting each other, by contrast the mechanism of long period variables (LPVs) was debated for a long time. In 1941 \citet{eddington1941} presented a last version of his theory of the Cepheids. Eddington thought the variation of luminosity resulted from a structure instability leading to periodic variations of radius and luminosity. This theory has since been widely used and applied in particular to Miras, Semi-regular, RR Lyrae.

After studying Miras stars, feeling growing doubts with the given explanation, and becoming aware of its inconsistencies, \citet{pba} suggested that Mira variability was due to the presence of a close companion around the star. 

Asteroseismology made significant progress since the CoRoT \citep{baglin} and \emph{Kepler} \citep{borucki} probes -- particularly on red giants : \citet{deridder,mosser2011,mosser2012a,mosser2012b,mosser2013,baudin,deheuvels,beck} and references therein. Rich power density spectra of the light curves are observed showing radial and non-radial modes providing a detection of the radial orders and angular degrees ($\ell=0, 1, 2, 3$) of the modes. Can these results be extended to the pulsation of LPVs whose characteristics: high amplitudes and presence of a single radial mode, are quite different than those of asteroseismology: low amplitudes, many simultaneous radial and non-radial modes giving a rich spectrum ?

The presence of companion was suspected around Asymptotic Giant Branch (AGB) stars to explain spiral-shaped nebulae following predictions by \citet{theuns} and \citet{mastrodemos}. Planetary nebulae with binary central stars are detected \citep[][and references therein]{jones}.  The issue of the presence of companions around Long Period Variables is then widely justified.

Finally, the pulsation paradigm being ruled out, one is led to wonder whether the presence of companions could explain the phenomenon of LPV.

The next section presents the main inconsistencies of the pulsation paradigm, section 3 is an analysis of Long Period Variables light curves. Section 4 focuses on modelling. In Section 5 the results are discussed before concluding in the last section.

\section{Substantial inconsistencies of the pulsation paradigm}

This research is motivated by the critical issues raised by the pulsation paradigm and not to offer a competing mechanism (the binarity hypothesis came afterwards to explain the variability given the inconsistencies of the pulsation paradigm). The most fundamental objection is the need to keep a perfect spherical symmetry to maintain a single radial pulsation mode. Any disturbance of the spherical symmetry giving rise to non radial modes.

\subsection{Pulsation and binarity}

Many recent studies seem to converge towards the presence of companions around many Miras or AGB stars:
\begin{itemize}
 \item \citet{kim} show evidence of a binary-induced spiral around the AGB star CIT6 previously suspected by \citet{dinh},
\item \citet{vanwinckel} found six binaries AGB stars with orbital periods ranging from 120 to 1800 days, 
\item \citet{mauron} found that half their sample of 22 AGB stars have elliptical emission that they understood binaries whose envelopes are shaped by a companion,
\item \citet{kervella2014,kervella2015} find a companion around L$_2$ Pup. 
\item \citet{maercker} show the presence of a companion around R Sculptoris to explain its large mass loss,
\item the situation is not so clear around R Fornacis: binarity is a possible hypothesis for \citet{paladini}, 
\item \citet{mayer2013} look for the signature of the companion interaction with the stellar wind of the symbiotic Mira star R Aquarii and the binary Mira W Aquilae,
\item \citet{boffin} find that FG Ser, a binary giant with a period of 630 days, is filling its Roche Lobe,
\item \citet{mayer2014} find a close companion around $\pi^1$ Gru with a period shorter than 10 years,
\item \citet{jeffers} detect an equatorial disc around IRC+10216, 
\item \citet{decin} a binary-induced shell around the same object.
\end{itemize}

By their presence, these companions break the spherical symmetry and the stars become unable to maintain a radial pulsation: the pulsation paradigm can no longer be applied. 

Moreover many Miras are thought to be symbiotic \citep{belczynski}, R Aqr has jets, is surrounded by an inner and an outer nebula and is believed to possess an accretion disc around its companion white dwarf \citep{nichols}. At least three characteristics of symbiotics are present: UV excess, H$\alpha$ emission, [OIII] emission. If these objects are symbiotics, we have to forget the pulsation paradigm. 

\subsection{Pulsation and convection}

\begin{figure}
  \centering
  \includegraphics[width=11cm]{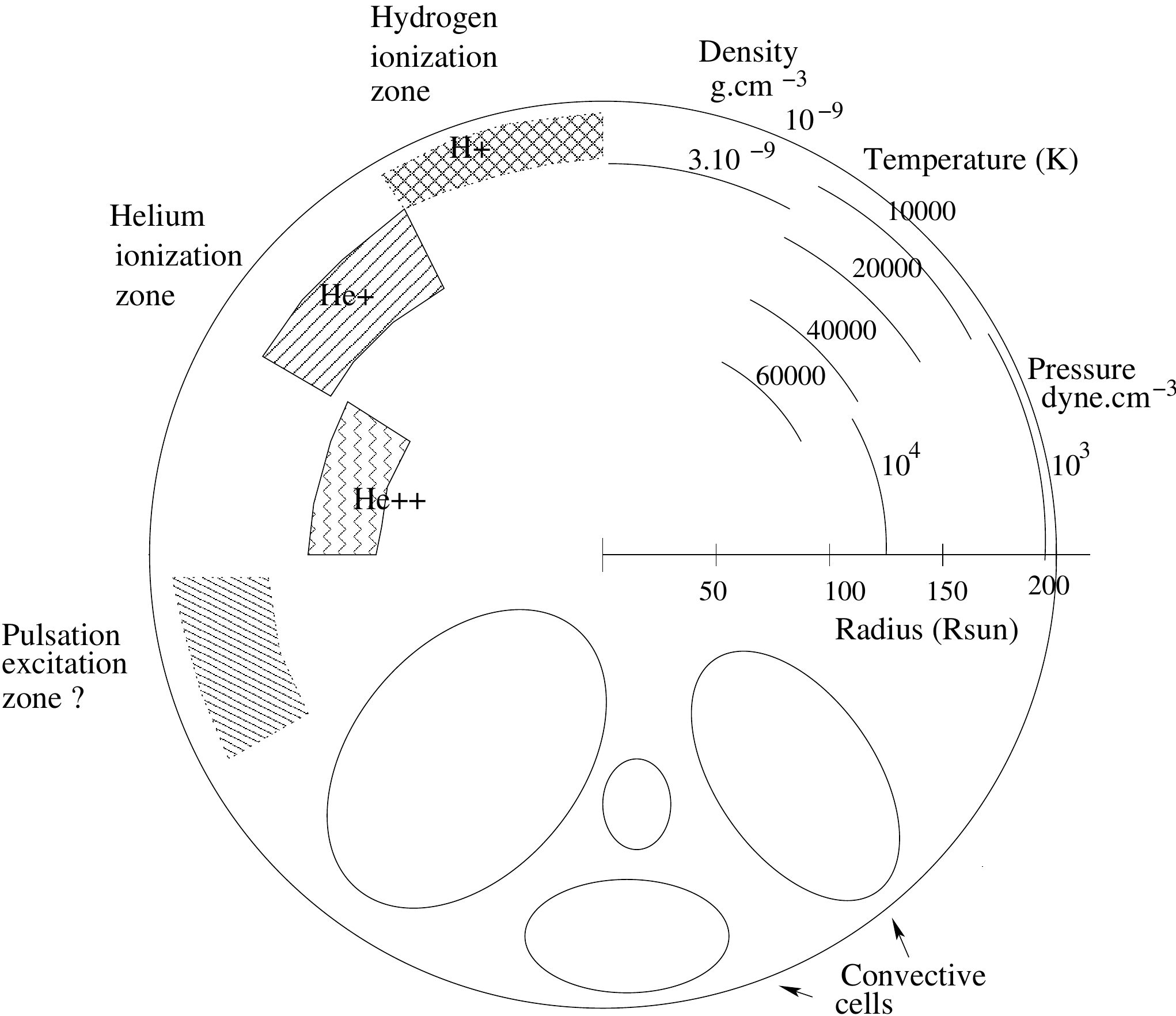}
  \caption{Sketch of an AGB star. The hydrogen or helium ionisation zones are supposed to be the source of the pulsation. Large convective cells are present throughout the envelope. The dense core at the centre of the star is not shown.}
  \label{FigStruct}
\end{figure}

Miras are ``cool'' stars, namely their effective temperatures is around 2500K. At these temperatures, many molecules are formed leading to a strong opacity. A deep convection occurs as the opacity in the main volume of the envelope is such that the heat produced in the core of the star cannot be dissipated by radiation: convection cells are formed and convey heat excess.
Convection is common in cool stars; calculations show that, concerning the Miras, convection cells are so large that a few only cover the whole surface of the star and convey more than 90 per cent of the energy. \citet{schwarzschild} estimated that convection cells are so large that only a few are apparent on the surface of red giants and supergiants. This estimation is confirmed by high-angular resolution observations \citep{cruzalebes}. Figure~\ref{FigStruct} gives a sketch of an AGB star interior showing the convective envelope and the zone which is supposed to be the source of the pulsation.

However pulsation and convection are two kinds of instability sharing the same cause linked to an inability of the star to dissipate energy by radiation alone. Yet, convection breaks the spherical symmetry: some cells move up while at the same time outside the cells, cooled matter is flowing down. But pulsation has to be radial (spherically symmetric). Pulsation and convection are therefore incompatible. If radial pulsation is present, convection has to be inhibited. Unable to explain how convection may be inhibited, \citet{zhevakin} supposed that as the star pulsates, then its convection is inhibited. Circular reasoning. Recent 3D simulations of red giants stars do show huge convection cells but no significant radial pulsations  \citep{woodward,freytag2008,brun}. Movies from these simulations show these huge convection cells\footnote{http://www.astro.uu.se/~bf/movie/dst35gm04n26/movie.html} consistent with observations.

So the pulsation paradigm faces two major difficulties: a) pulsation is deprived of its driving mechanism by convection, and b) convection breaks the spherical symmetry which then loses the ability to maintain a radial pulsation.

\subsection{The pulsation mechanism}
The mechanism proposed, called $\kappa$-mechanism \citep{baker} is based upon the variation of the opacity with temperature. From this point of view, the opacity of the external layers rises with temperature, generating instability: when the temperature rises, the opacity rises also and then energy transfer is less efficient. All the produced energy cannot be dissipated and the temperature continues to rise. This increase in temperature inflates the star which then cools down and deflates before starting a new cycle.

In the case of the Cepheids the suspected zone driving the $\kappa$-mechanism is the helium ionisation zone \citep{xiong2007}. For the Miras the zone responsible for instability has not been firmly established, some authors suspect the hydrogen ionisation zone. Anyway this mechanism raises a problem of synchronisation. In order to explain the light curves by a pulsation, the pulsation mode should be radial: pulsation must spread at the same time on the whole surface of the star, or in other words, must keep a spherical symmetry. 

But how to explain a spherical symmetry when the instability is not produced at the centre of the star? How parts located far from each other (for example at the antipodes) could remain synchronised when they are beyond a pulsation wavelength? The time needed to exchange information between two parts far from each other is longer than the pulsation period. No synchronisation is then possible.

\subsection{Solar--like and Mira--like oscillations}
Recent advances show clear differences between the solar--like oscillations and the Mira--like variations phenomena. The red giants solar--like oscillations have very low amplitudes of the order of 0.0001 mag, so they are invisible on the light curves, Miras amplitudes have several magnitudes. \citet{lebzelter2005} consider very unlikely that a stochastic excitation mechanism produces such amplitudes in the fundamental mode. In addition, in the case of solar--like oscillations, several modes of different spherical and radial orders are excited simultaneously resulting in a rich spectrum showing several peaks. On the contrary, changes in Miras should be assigned to a single radial mode ($\ell=0$) of low radial orders: fundamental or first or second harmonic. The evolution from Red Giants oscillation modes to Semi--Regulars ones does not show a drastic increase of amplitudes or a tendency to keep only one radial mode \citep{banyai}.

This observation comes back to a very fundamental issue: why Miras would preserve only one pulsation mode. The superposition principle is a fundamental principle widely observed for all linear systems. According to this principle the movements of an oscillating system is the result of the superposition of different normal modes of the system. Why Miras, unlike other red giants, do would develop only one radial mode with a very high amplitude? No reply to this objection was given, yet it should, by itself, justify the rejection of the pulsation paradigm.

\section{Light curves analysis}

Long Period Variables are, as their name suggests, mainly characterised by their light curves. So it is natural to focus on the light curves and try to maximise the learning potential of these data.

In the following we compare the characteristics of the observed light curves with the predictions of the pulsation paradigm and of the presence of a companion.

The traditional approach for period determination is based upon the \emph{O--C method} (i.e. observed minus calculated) which follows, for a given period, the difference between the observed and the calculated dates of maximum. This method presents some drawbacks:
\begin{itemize}
\item Only few data are used: dates of observed maximum (or minimum),
\item Maximum dates are not precise, due to the lack of observations close to the maximum and to the intrinsic variations of the variable from cycle to cycle, and the limited precision of the visual magnitudes,
\item Some objects cannot be observed all along the year and some maximum may be not observed for some cycles.
\end{itemize}

Today, Fourier or period-time analysis make use of the full set of data revealing the maximum possible information \citep{szatmary}.

\subsection {Data}

\begin{figure}
  \centering
  \includegraphics[width=10cm]{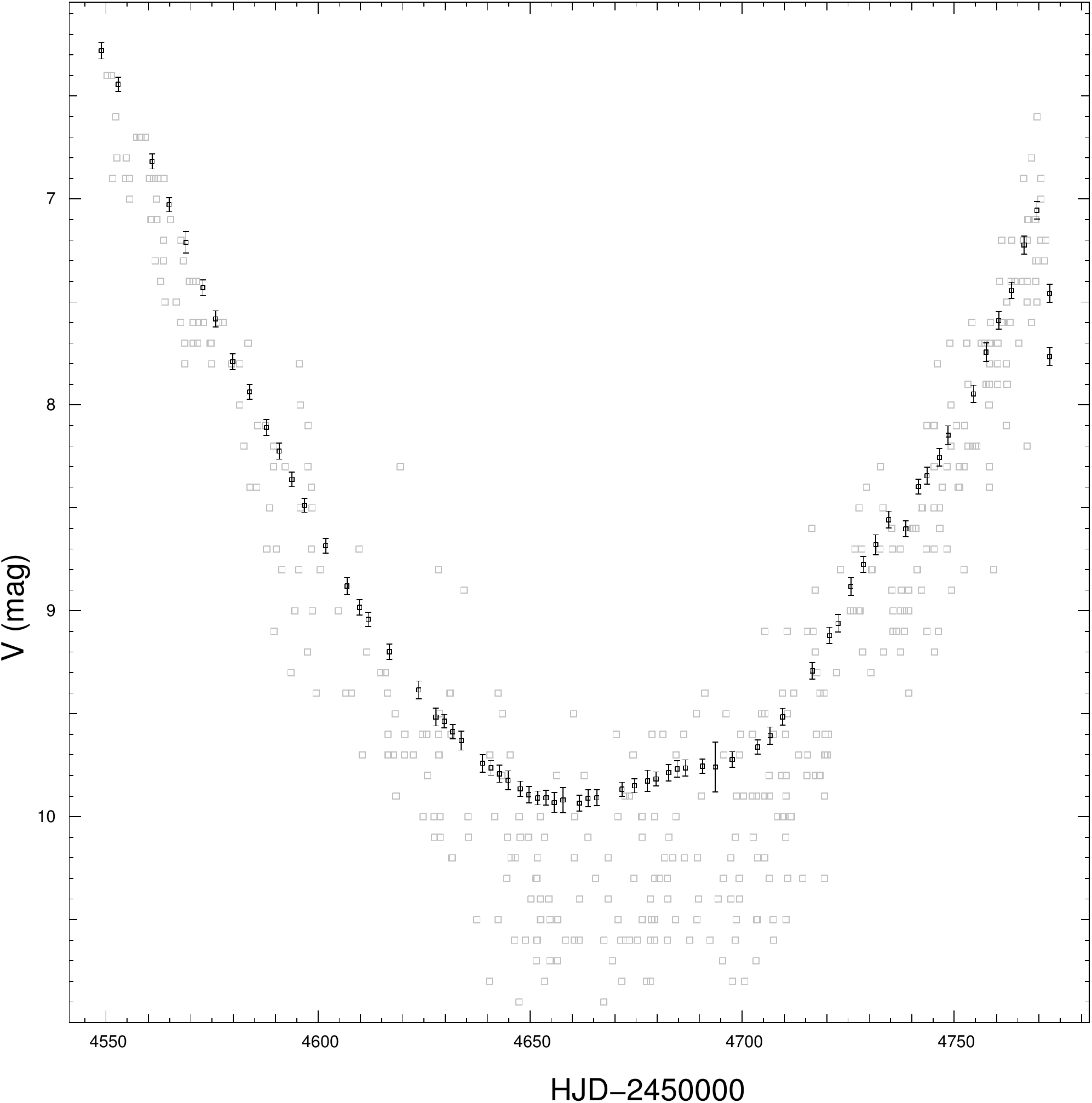}
  \includegraphics[width=10cm]{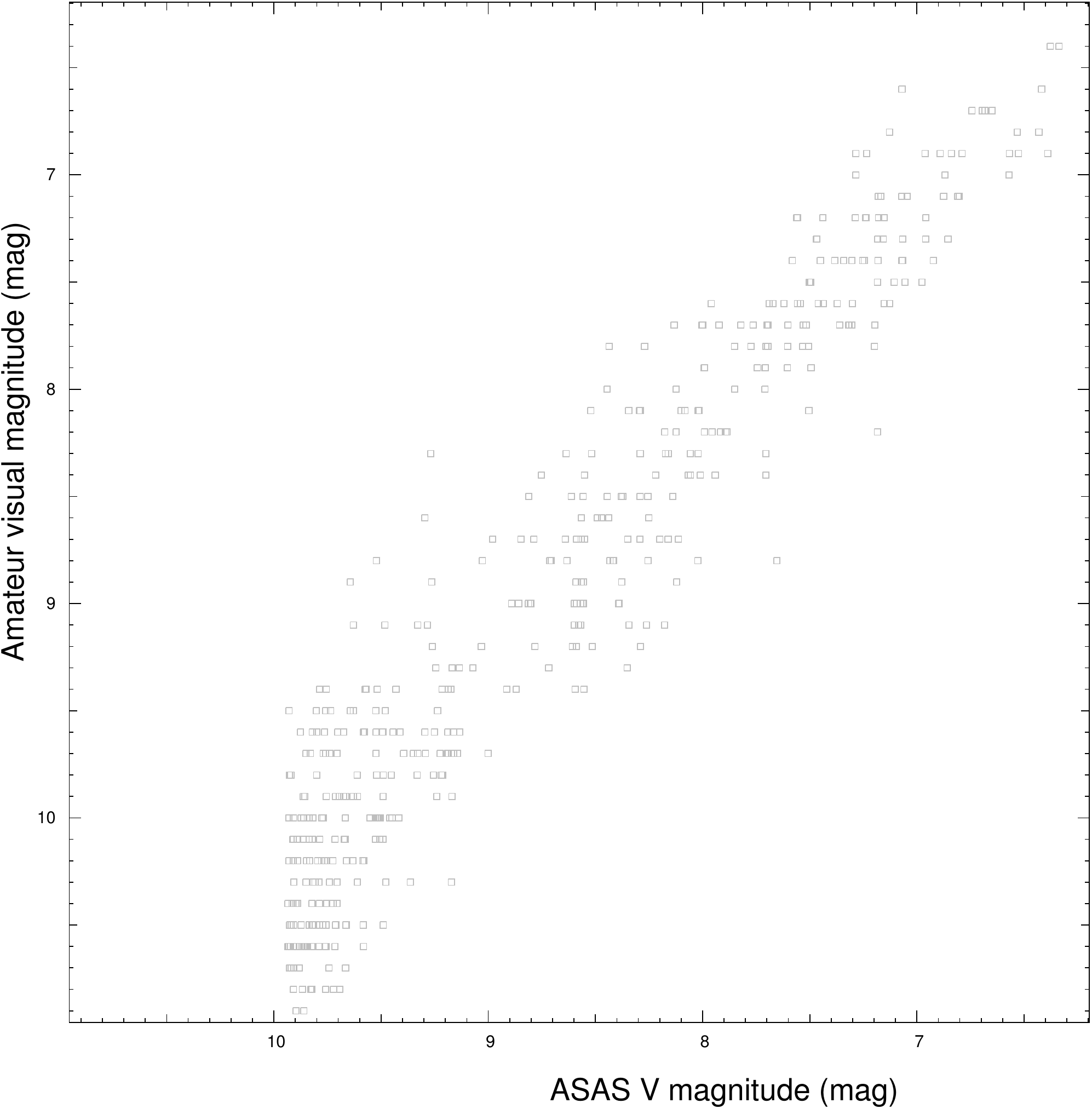}
  \caption{Left panel : part of R Aql light curve with visual data (Gray squares) and ASAS data (black with error bars). Right panel : Visual magnitudes from amateur databases (AFOEV and AAVSO) versus ASAS $V$-band magnitudes at the same dates. Amateur error is estimated to 0.35 mag.}
  \label{FigASAS}
\end{figure}

\subsubsection{Amateur data} 
Patient magnitude recording of numerous variables for more than a century, mainly from amateurs, has made it possible to draw detailed light curves of these variables and is therefore the main source of information. Some variable stars observers associations open their databases for thousands of variables.

We used photometric data from the Association Fran\c{c}aise des Observateurs d'Etoiles Variables\footnote{http://cdsarc.u-strasbg.fr/afoev/} (AFOEV), and the American Association of Variable Star Observers\footnote{http://www.aavso.org/} (AAVSO). 

The low level of accuracy of these data is to a large extent compensated by the duration of the observation campaign and the number of observations - in some cases they go back to 1840 and put forward 70\,000 individual measures, their even distribution in time with the exception of the two World War periods. Professional observations are far more precise, and on a wider wavelength range, but they can't cover such an extended period of time, or so many objects.

These amateur data provide an opportunity to precisely determine Miras periods, their variations with time, and also their amplitudes and variations, and more generally their shapes that are close to a sine curve or showing bumps or double humps; evidence of a possible evolution of these shapes over time can also be put forward.

For each object a text file gives the dates of observation, the magnitudes observed (in most cases visually), the name of the observer. No standard deviation is given for visual observations, we estimated it in comparison with professional data (see below).

\subsubsection{ASAS} The All Sky Automated Survey\footnote{http://www.astrouw.edu.pl/asas/} (ASAS) database is available with a catalogue of more than $18\,000$ regular variables \citep{pojmanski}. $V$-band magnitudes are given with their associated errors ($\approx 0.05$ mag). The covered periods are 1998-2000 (ASAS-2) and 2000-2009 (ASAS-3).

Figure~\ref{FigASAS} shows a part of R Aql light curve with data drawn from amateur and ASAS databases allowing to estimate the amateur error. Comparing these data with the magnitudes recorded by the amateurs, we obtained an estimation of the standard deviation of the amateur data of 0.35 mag. 

\subsubsection{Infrared} The infrared photometry of \citet{whitelock2000,whitelock2006} allows to compare the visible and infrared light curves of many objects. \citet{whitelock2000} present the near infrared, \emph{JHKL}, photometry of 193 Mira and Semi--Regular variables which were observed by Hipparcos, and \citet{whitelock2006} of 239 Galactic C-rich variable stars. The period covered depends on the objects between 1975 and 2000. The photometry is accurate to $\approx0.03$ mag at \emph{JHK}, and $\approx 0.05$ mag at \emph{L}.

\subsubsection{GCVS} The General Catalogue of Variable Stars\footnote{http://www.sai.msu.su/gcvs/gcvs/} (hereafter GCVS4) lists all bright variable stars giving their type (Mira, Semi--Regular, Cepheid, \dots), periods, maximum and minimum magnitudes, dates of a maximum. GCVS4 defines Miras according to their periods (between 80 and 1000 days) and their amplitudes (from 2.5 to 11 mag). Semi--Regulars differ from Miras only by their smaller amplitudes (SRa), or their irregular light curves (SRb, SRc and SRd). This distinction between Miras and Semi--Regulars is quite arbitrary: the periods and amplitudes distributions don't show two separate populations.

\subsubsection{CoRoT}
The CoRoT space mission \citep{auvergne} provided high photometric data along successive runs of 25 and 150 days from 2007 to 2012. \citet{lebzelter2011} used data from four long runs to search for LPVs and selected 52 candidates. The author notes a few systematic shifts in the light curves of the order of 0.02 mag. The data are available through the CoRoT Public Archive\footnote{http://idoc-corotn2-public.ias.u-psud.fr/}.

\subsubsection{\emph{Kepler}} 
The \emph{Kepler} space mission provided four years of continuous high photometric quality data. \citet{banyai} study the variability of M giants stars based on these data. Due to the telescope roll every quarter of a year, the light curves show jumps, moreover an unexplained \emph{Kepler-year} signal is present in the data with an amplitude of about 1 percent. The authors emphasise the need for caution when using \emph{Kepler} data for investigating long-term phenomena that needs data over hundreds of days. Data are publicly available\footnote{http://archive.stsci.edu/kepler/}.

\subsection {Light curve fitting} 

The simplest way is to fit a constant period sine curve on the light curve. Starting with the GCVS4 given period, we proceed according to two kinds of fits:
\begin{enumerate}
\item The light curve is defined with the help of six parameters, namely:
  \[
  \begin{array}{lp{0.8\linewidth}}
    t_\mathrm{max}  & date of a maximum           \\
    P             & period                      \\
    m_\mathrm{max}  & magnitude of maximum       \\
    m_\mathrm{min}  & magnitude of minimum        \\
    f             & asymmetry factor            \\
    P'            & period variation with time  \\
  \end{array}
  \]
\item The light curve is defined by two parameters: period and its variation with time. The fit is performed on the mean light curve derived from these two parameters.
\end{enumerate}
In each case, a non-linear least-squares Levenberg-Marquardt method is applied.

The advantage of these methods is obvious: we obtain not only the period and its variation with time but also the mean curve, the mean maximum and minimum magnitudes, the asymmetry factor and a date of maximum.

The phase $\varphi$ is deduced from a maximum date and the period function of time $t$ by:
\begin{displaymath} 
  \varphi = 0\; \mathrm{ for}\; t=t_\mathrm{max}
\end{displaymath}
If $P$ is constant with time:
\begin{displaymath}
  \varphi(t) = \frac{t-t_\mathrm{max}}{P_\mathrm0}\,\mathrm{modulo}\, 1
\end{displaymath} 
Otherwise $P$ may be approximated by a linear function of time with $P_\mathrm0$ the period at a given date of maximum $t_\mathrm{max}$:
\begin{displaymath}
  P(t) = P_\mathrm0 + P'(t-t_\mathrm{max})
\end{displaymath} 
But in some cases period variations are not linear, and $P$ is expressed by a linear interpolation between an arbitrary number of couples of dates and periods. Phase is obtained by numerical integration:
\begin{displaymath}
  \varphi(t) = \int\limits_{t_\mathrm{max}}^{t}{\frac{1}{P(t)}dt}\,\,\mathrm{modulo}\,\, 1
\end{displaymath}
or for a linear variation of $P$ $(P'\neq0)$:
\begin{equation}
  \varphi(t) = \frac{1}{P'} \log \left(1+\frac{P'}
 	 {P_\mathrm{0}}(t-t_\mathrm{max})\right)\,\mathrm{\ modulo}\ 1 \label{phase}
\end{equation}

When using the asymmetry factor $f$, the synthetic light curve is made of two half-sine curves, the first half from maximum to $(1-f)$ phase:
\begin{equation}
  m = \frac{m_\mathrm{max}-m_\mathrm{min}}{2}
  \cos\left(\frac{\pi\varphi}{1-f}\right)+
  \frac{m_\mathrm{max}+m_\mathrm{min}}{2}\label{mag1}
\end{equation}
and the second one between $(1-f)$ phase and the following maximum:
\begin{equation}
  m = \frac{m_\mathrm{max}-m_\mathrm{min}}{2}
  \cos\left(\pi\frac{\varphi-1}{f}\right)+
  \frac{m_\mathrm{max}+m_\mathrm{min}}{2}
  \label{mag2}
\end{equation}

This synthetic light curve generally gives a good fitting when the asymmetry factor is not too low ($f>0.2$) but is not well suited for low asymmetry factors. 
The process we used for most objects is the following:
\begin{enumerate}
\item The parameters starting values are given by GCVS4,
\item The parameters to fit are first limited to the period and a date of maximum, the remaining four parameters being fixed,
\item Then extended to asymmetry factor, maximum and minimum magnitudes,
\item If discrepancies appear visually, the fit is allowed to vary the period variation, and the procedure is eventually adapted to obtain a better fitting.
\item A fit of the mean curve is then performed with the two starting parameters (period and period variation) just obtained.
\end{enumerate}

Thus the method is almost automatic. This process is used to avoid a ``derailment'', each step being visually checked. In some cases, when the GCVS4 data are too far from the observations (or when GCVS4 gives no period), starting parameters are manually modified. In some other cases, there are long time intervals without data and the fit has many minimums depending on how many periods are included in the time interval, a human intervention is then necessary.

In some cases, no satisfactory fitting was obtained without splitting data according to the period variations: these objects showing repetitive or sudden period variations.
\begin{figure}
  \centering
  \includegraphics[width=11cm]{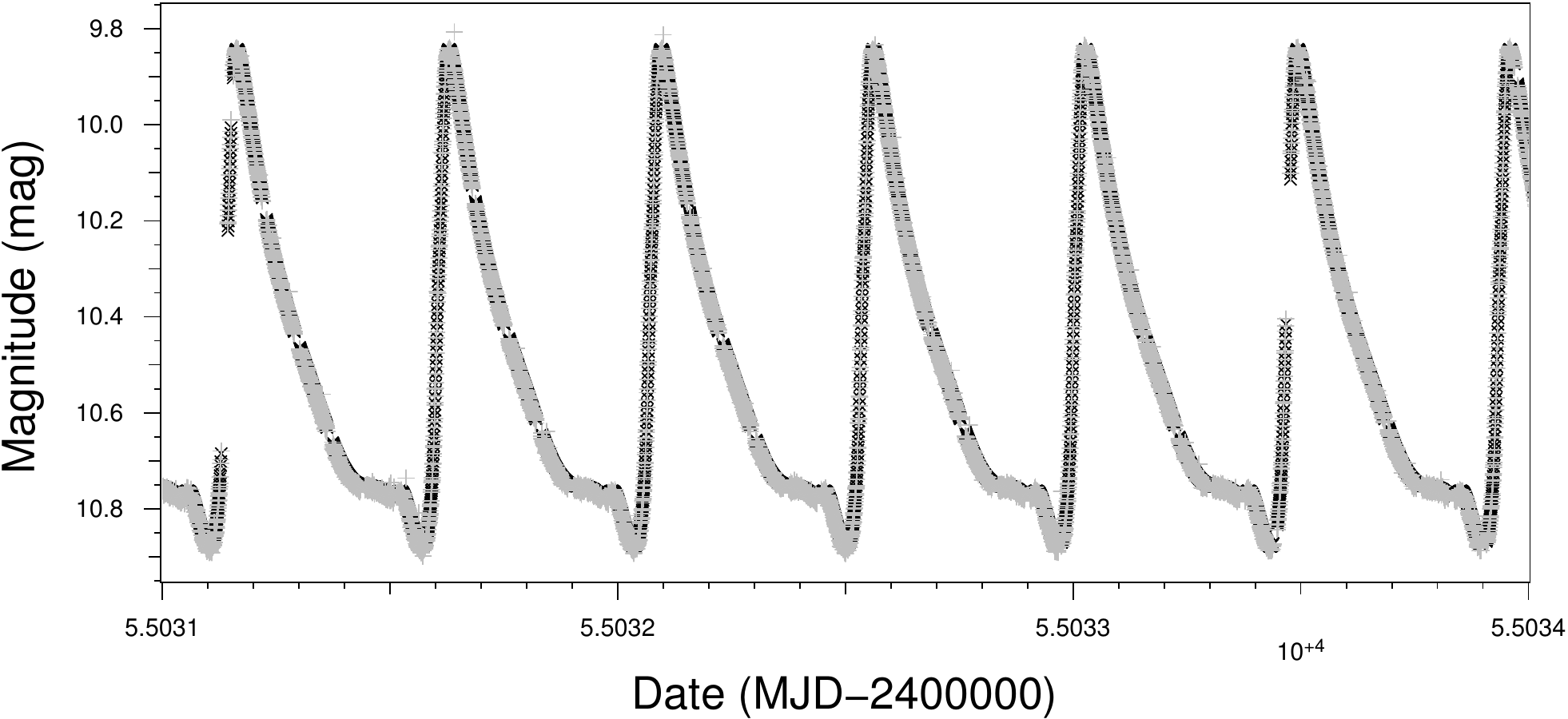}
  \caption{Part of the light curve of the RR Lyrae CoRoT 0103800818 (Gray plusses) and mean light curve (black crosses) at the same dates. Although the mean curve is calculated on the whole data set, the differences between the light curve and its mean is very small.}
  \label{FigRR}
\end{figure}

To check our approach, we tested it with CoRoT data to take advantage of the precision photometry. We chose a RR Lyrae because the number of periods of LPV variables covered by CoRoT is not sufficient. The RR Lyrae variable CoRoT~$0103800818$ is reported by \citet{szabo}. We first fitted two half sine waves to get the period and then we fitted the mean light curve. Figure~\ref{FigRR} shows this mean light curve is  very precisely adjusted on the data. The constant period is 0.46597 days.

Table~\ref{tabfit1} gives the fitting results for 163 Miras and Semi--Regulars (SRa) with sufficient data and showing fixed periods or with a constant variation period rate.

Table~\ref{tabfit2} gives the fitting results for 116 Miras and Semi--Regulars with sufficient data and no monotonously varying period. A period is associated with each given date. Between two successive given dates, the period is supposed to vary linearly with time. Up to six dates (and their associated periods) are given.

Table~\ref{tabfit3} gives the fitting results for 28 Semi--Regulars (SRb) with sufficient data and showing clear periods.

\subsubsection {Splines fits}
 One difficulty of light curve analysis is the time unevenness of observations. In order to avoid this drawback, we tried to fit a spline on the data rather than interpolating, the results were not improved and we dropped this solution. 
\subsubsection {Mean light curves}
With the help of the fitted period, the phase is calculated for each observation date, and the mean light curve is obtained.

\subsubsection {Period--luminosity relation}
The Miras are among the most luminous objects in the infrared, therefore it is very important to get a period--luminosity relation. \citet{whitelock2008} derived such relations giving the absolute $K$ magnitude $M_K$ from the period $P$ in days:
\begin{displaymath}
M_K = -3.69\,\log_{10}P-7.25
\end{displaymath}

\citet{glass} obtained from MACHO data for Miras (their sequence C) in the LMC:
\begin{displaymath}
M_{K_S} = -3.56\,\log_{10}P+ \mathrm{constant}
\end{displaymath}
It must be stressed that these relations are based upon observations not on theoretical grounds.

Such a relation can be estimated for a binary. According to the third Kepler law, the period $P$ is related to the distance $a$ between the two bodies centres of mass and the total mass $M$ of the system:
\begin{displaymath}
  \frac{a^3}{P^2}=\frac{\mathrm{G}M}{4\pi^2}
\end{displaymath}
where G is the gravitational constant.
The luminosity $L$ of a star is given by: 
\begin{displaymath}
  L=4\pi R^2\mathrm{\sigma} T_\mathrm{eff}^4
\end{displaymath}
where $\mathrm{\sigma}$ is the Stefan constant, $T_\mathrm{eff}$ the effective temperature of the star, and $R$ its radius. Assuming the radius is the distance $a$ of the two bodies (i.e. the mass ratio is sufficiently different from one):
\begin{displaymath}
  L=4\pi \left(\frac{P^2\mathrm{G}M}{4\pi^2}\right)^{2/3}\sigma T_\mathrm{eff}^4
\end{displaymath}
for a given effective temperature and mass, we obtain an estimation of the slope of the period--luminosity relation:
\begin{displaymath}
M_K = -3.33\,\log_{10}P + \mathrm{constant}
\end{displaymath}
\noindent this raw evaluation is close to the slope of the above period--relation of \citet{whitelock2008} $-3.69$ or \citet{glass} $-3.56 \pm 0.29$.

The companion hypothesis seems compatible with the observed period--luminosity relation. 

\begin{figure}
  \centering
  \includegraphics[width=10cm]{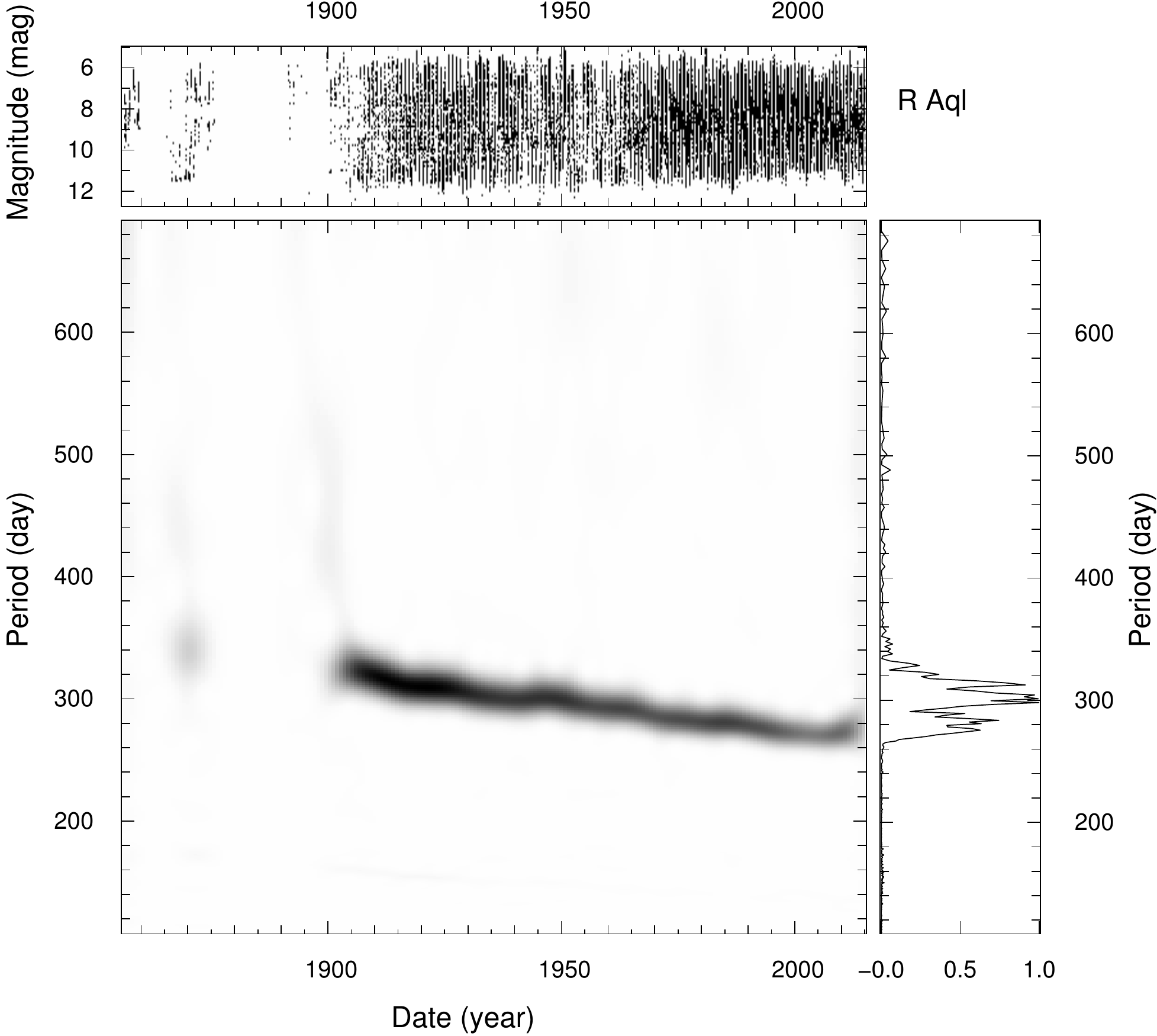}
  \bigskip
  \includegraphics[width=10cm]{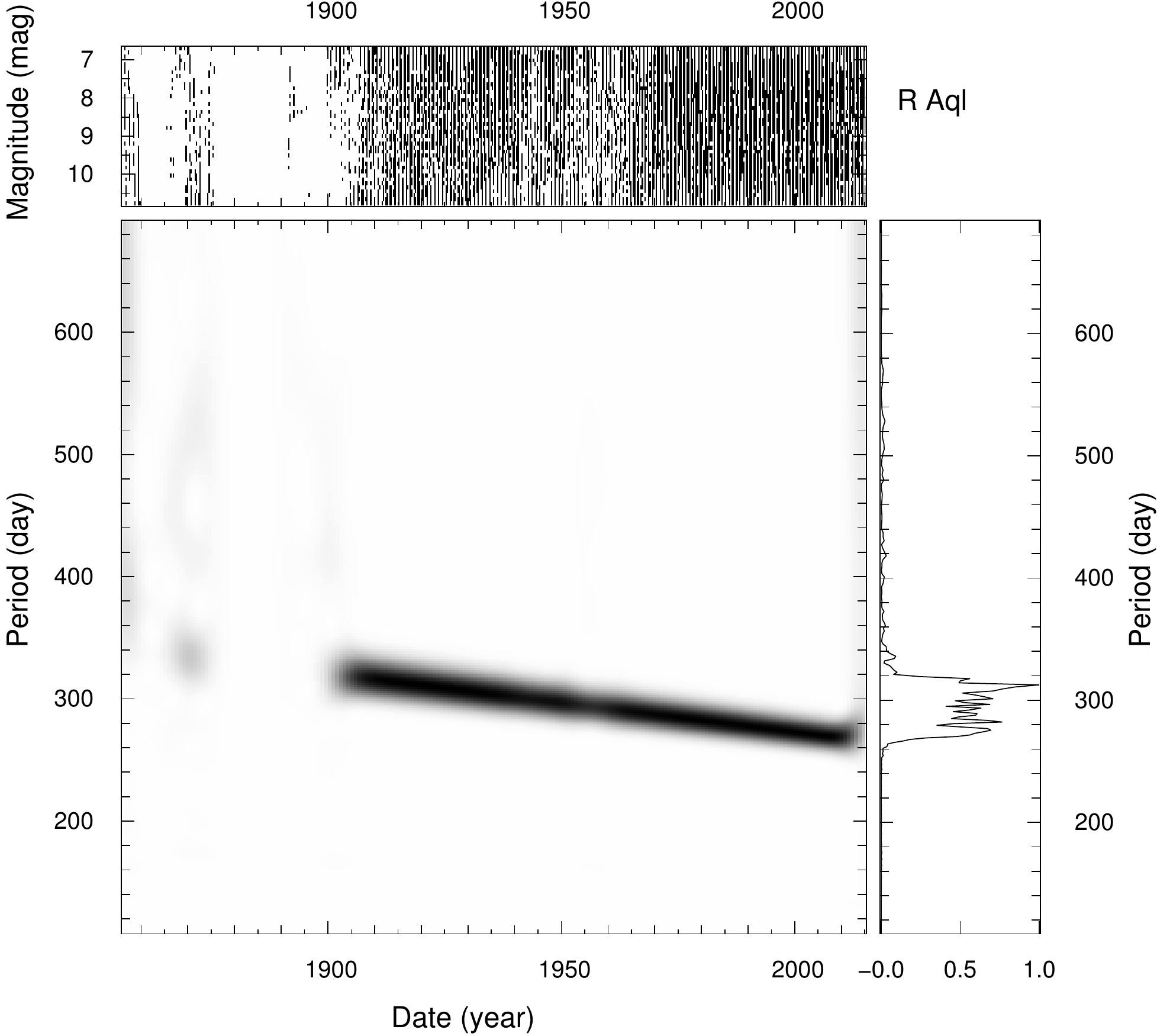}
  \caption{Time Analysis of R Aql observed and simulated light curves. Top. R Aql observed data. Top panel: light curve from amateur databases. Central panel: time--period map. Right panel: power spectrum on the whole data set (scaled to maximum) : the period of the variable is regularly decreasing. Bottom. R Aql simulated data: magnitudes estimated at the dates of the observed ones with a fixed amplitude and a period decrease of $0.48$ day/year.}
  \label{FigRAqlO-C}
\end{figure}

\begin{figure}
  \centering
  \includegraphics[width=11cm]{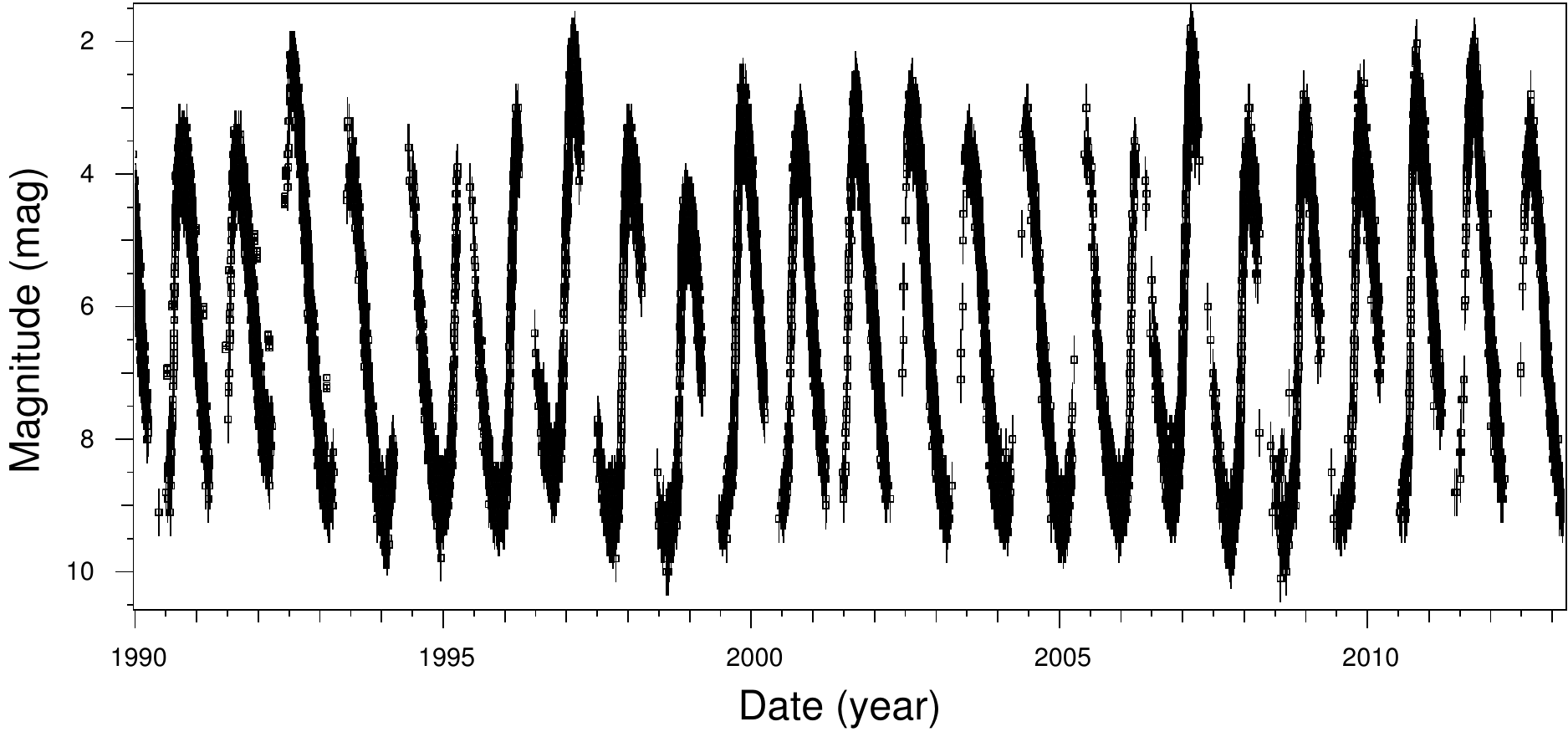}
  \bigskip
  \includegraphics[width=11cm]{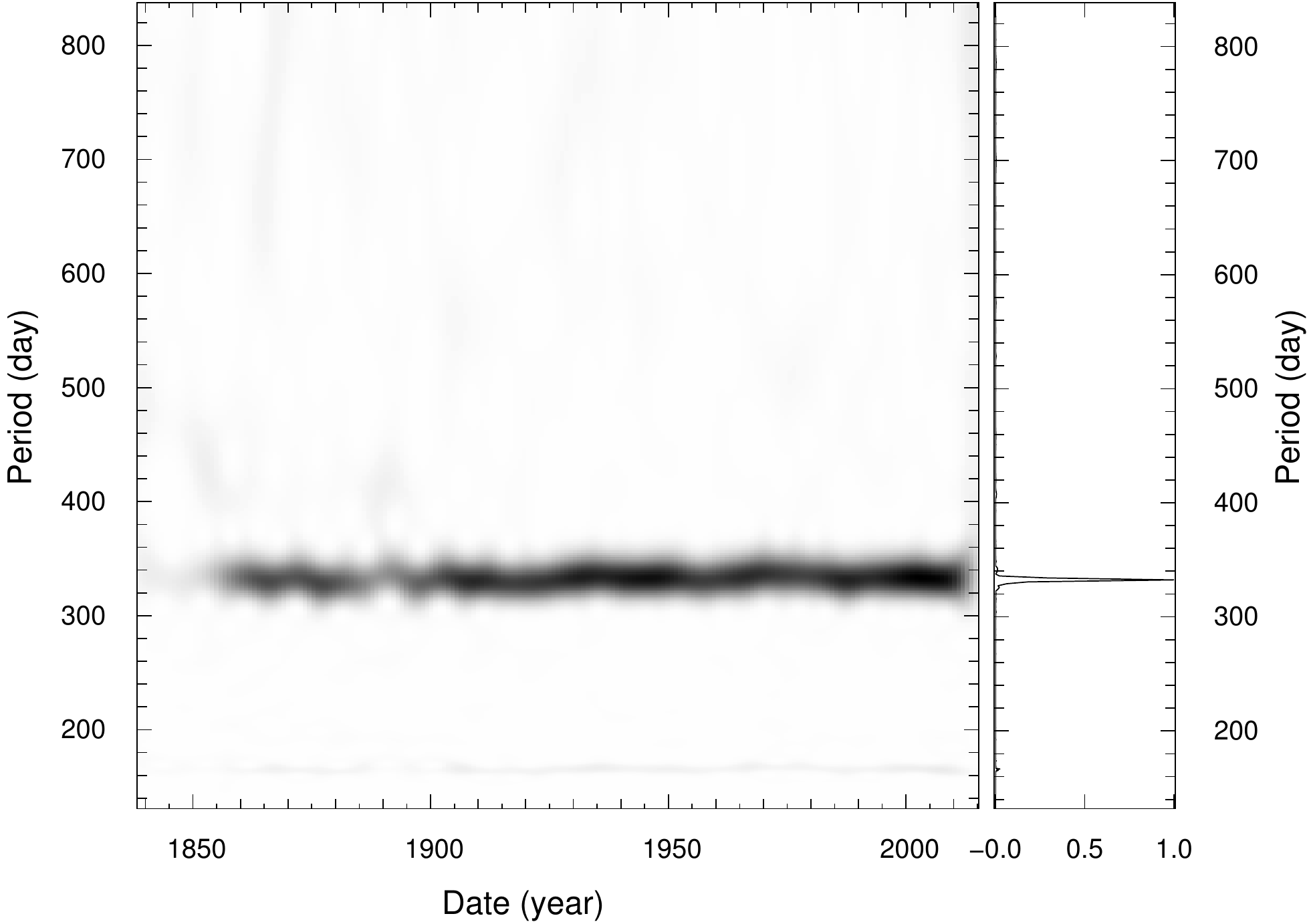}
  \caption{Mira (o Cet) Light Curve Time Analysis drawn with the help of AAVSO and AFOEV data. Top. Light curve of Mira from 1990. Bottom. left panel: time--period map from 1850 to present, the period does not vary with time, right panel: power spectrum on the whole data set (scaled to maximum).}
  \label{FigMira}
\end{figure}

\subsection {Power spectrum of the light curves}

Using CoRoT or \emph{Kepler} data many authors \citep{deridder,mosser2011,mosser2012a,mosser2012b,mosser2013,deheuvels,beck} showed that red giants present simultaneous radial and non-radial (dipole and quadrupole) modes of oscillation of different radial orders. However \citet{baudin} identify only radial modes on HR2582.

The question is whether there is a continuity between these solar--like modes of oscillation characterised by many simultaneous modes, low amplitudes and short lifetime and the behaviour of Miras with a single period, very large amplitude, and long lifetime beyond century. In this regard, the study of Semi--Regular shed interesting light if their behaviour is intermediate between that of red giants and Miras.

\citet{banyai} analyse the variability of both RGB and AGB M giant stars using \emph{Kepler} photometry. In many cases, after reconnecting the light curves from different quarters, they found only two significant period in the light curves, but for the lower amplitude stars over fifty. Drawing a Petersen diagram (ratio of periods versus the longer period), they find no structure for the low amplitudes nor for the Miras, only a clump for a period ratio of $\approx 0.7-0.8$ and periods below 100 days. Anyway, they suggest an indication of a transition between the two types of excitation (solar--like and Mira--like) around a period of 10 days ($1.2\, \mu\mathrm{Hz}$). 

The use of \emph{Kepler} and OGLE data does not allow \citet{mosser2013} to distinguish between AGB and RGB based on the power spectra of their light curves. They attempt an extrapolation of oscillation modes identified from the spectra with the highest frequencies to the spectra with very low frequencies. After this identification, they found a gradual disappearance of the non-radial modes (dipole and quadrupole) with decreasing frequency. They conclude that semi--regularity is due to the small number of stochastically excited oscillation modes that are observed. However, they believe that the variability of Miras cannot be explained by solar--like oscillations in particular because of the observed amplitudes far too important. \citet{lebzelter2005} reach the same conclusion.

The study of \citet{stello} shows the presence of radial and non-radial modes in Semi--Regular variables, the power oscillations in the dipole modes increasing relative to that of radial modes when the object is brighter. Quite the contrary \citet{mosser2013} note a decrease of the non-radial mode both for dipole and quadrupole modes relative to radial modes.

\citet{hartig} show that all the Semi--Regular variables they study have multiple modes. However, they note that non-radial oscillations of the size required for M giant semi--regular variability would result in unrealistic distortions. 

\subsection {Time--period analysis}

The first step of time--period analysis is to obtain uniformly spaced data by interpolation of data or splines. Many power spectrum are then derived from the data multiplied by the profile of a sliding cubic B-spline centred at zero on a uniform grid, namely :

\begin{displaymath}
\left\{\begin{array} {cc}
  \left(|x|/2 - 1\right)x^2 +2/3 & |x|\leqslant1\\
  \left(2-|x|\right)^3/6 & 1<|x|<2\\
   0 & |x|\geqslant2
\end{array} \right.
\end{displaymath}

The full width at half maximum of this cubic B-spline is $1.4447$ in unit of $x$.

To verify the efficiency of our time--period analysis, we compared the results obtained with real data and with synthetic data derived from our fitting parameters (equations (\ref{phase}), (\ref{mag1}), (\ref{mag2})) at the same dates than the observed ones (Fig.~\ref{FigRAqlO-C}). Synthetic data are obtained with constant maximum and minimum magnitudes and a linear variation of period as fitted on the real data. Comparison shows very similar results.

The first observation is that the light curves of most Miras show stable periods with time. Fig.~\ref{FigMira} shows a time--period map of the light curve of Mira. The 332 days period remained stable over the last 160 years ! Code's argument against Hoyle and Lyttleton model stating these objects are aperiodic loses its validity.

\begin{figure}
  \centering
  \includegraphics[width=11cm]{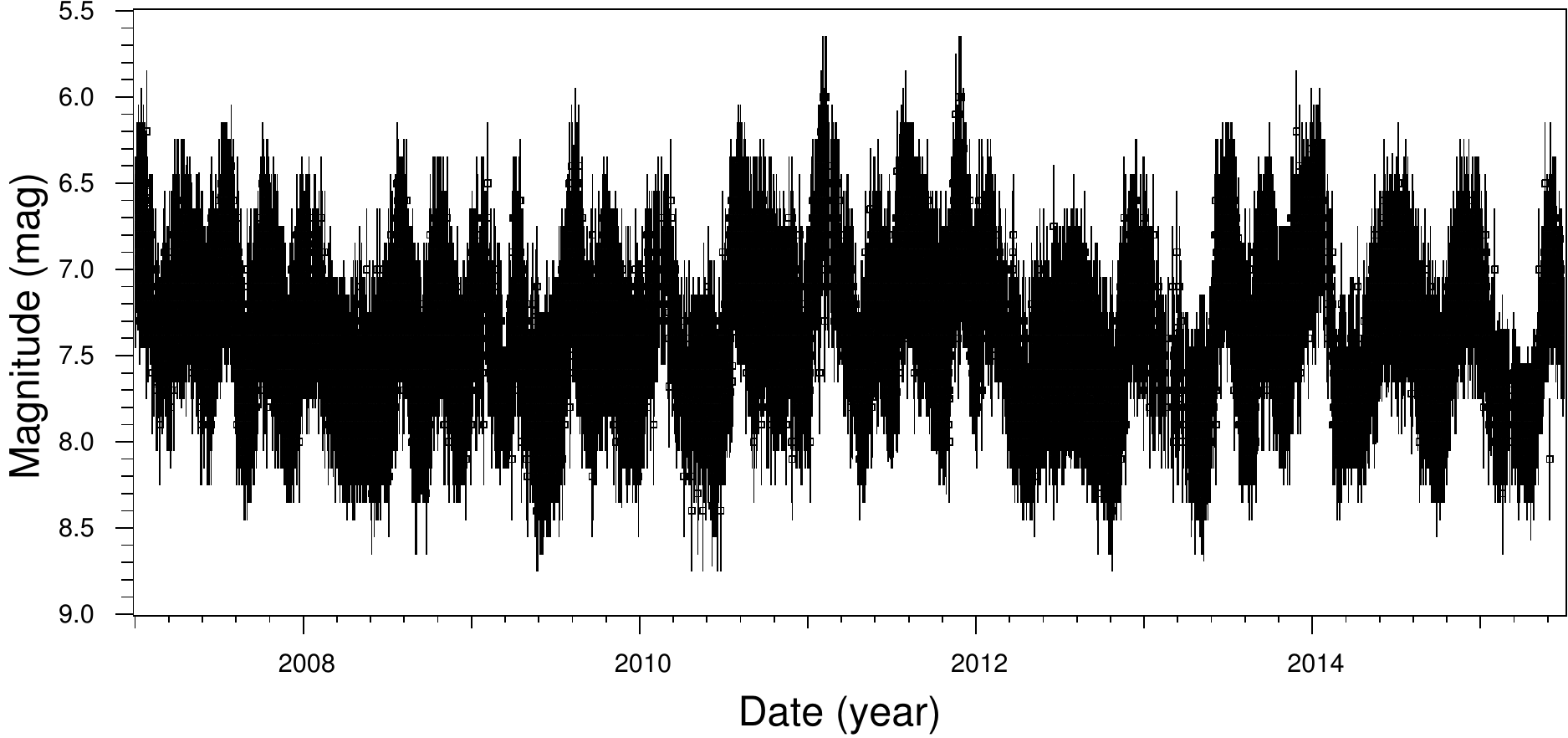}
  \bigskip
  \includegraphics[width=11cm]{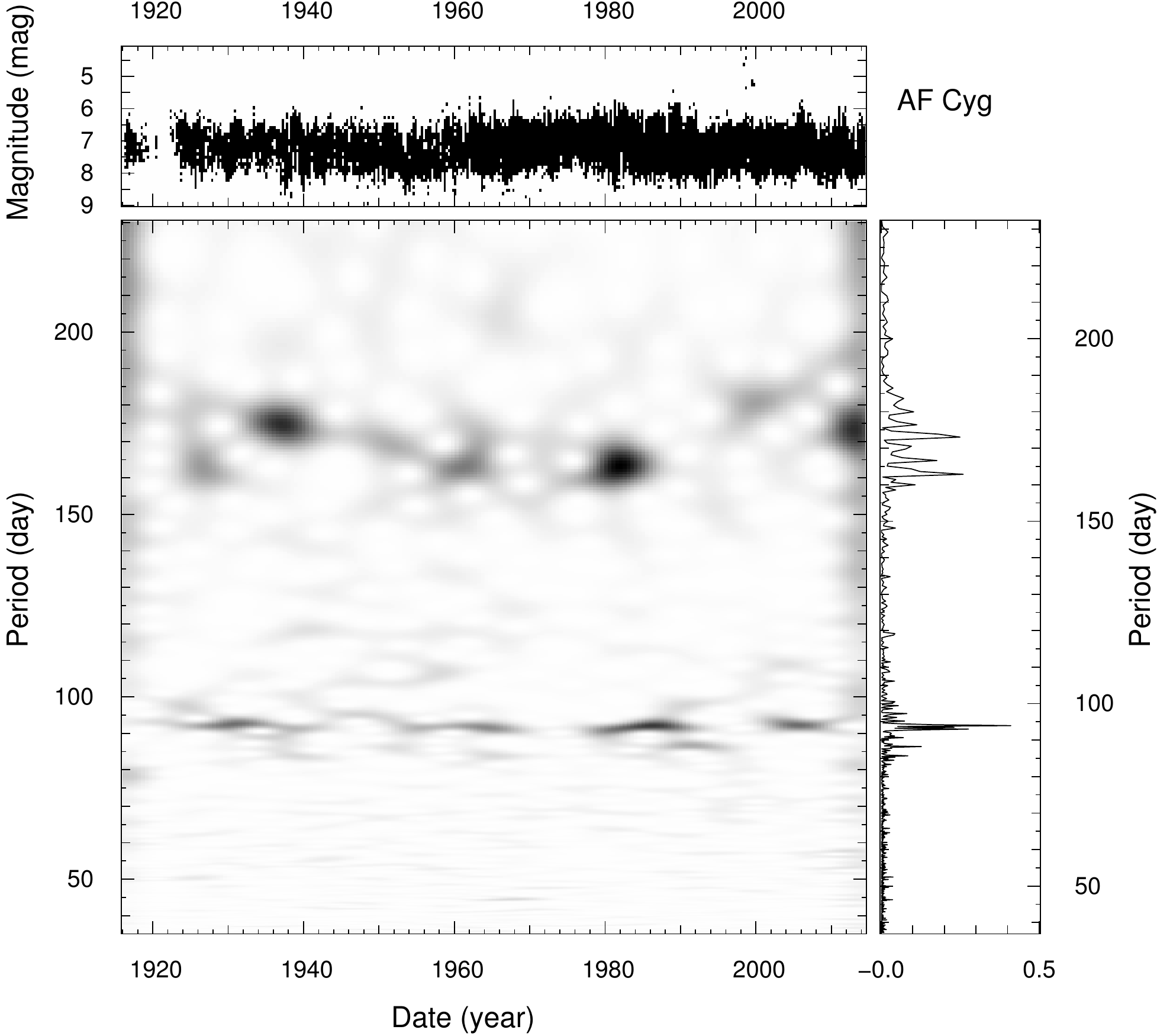}
  \caption{Time Analysis of AF Cyg light curve. Top: Part of light curve from amateur databases. Bottom panel: time--period map. Right panel: power spectrum on the whole date set (scaled to maximum): this Semi--Regular has a nearly constant period of 92.9 days.}
  \label{FigAFCyg}
\end{figure}

\citet{hartig} find that the periods derived for Semi--Regular variables at a specific time interval are not necessarily the same for observations at a different time.

In this regard, \citet{banyai} show wavelet maps of the Semi--Regular AF Cyg (their figure~9) based upon the four years \emph{Kepler}'s data set and upon the AAVSO data at the same dates: the two maps are very similar. Figure~\ref{FigAFCyg} shows our time--period analysis of the whole secular AFOEV and AAVSO data set for the same object. No period is evident on the light curve, nevertheless a main period of 92.9 days ($0.125\,\mu\mathrm{Hz}$) is visible on the time--period map and on the power spectrum. Fitting a sine wave gives a non-significant period variation of 0.005 day by year. However when this analysis is limited to the \emph{Kepler} window from 2009 to 2013 this period disappears and a feature around 175 days ($0.066\,\mu\mathrm{Hz}$) is prominent as shown in their figure. 

Thus we emphasise that a long data set is preferable to study the long-term behaviour of LPV although photometry is not as accurate. Even in the case of the Semi--Regular AF Cyg a main period of 92.9 days remains stable over time. More precisely, the power spectrum of AF Cyg has between 155 and 195 days ($0.059$--$0.075\,\mu\mathrm{Hz}$) several fairly large peaks and a narrow 92.9-day peak ($0.125\,\mu\mathrm{Hz}$). In the time-frequency diagram (Fig.~\ref{FigAFCyg}) the peaks around 175 days show short lifetimes (as confirmed by their FWHM), the narrow 92.9-day peak appears throughout the century with more or less periodic disappearances.

If some frequencies can be explained by oscillations, it seems clear, as in the case of AF Cyg, that next to a rich spectrum which can be attributed to radial and non-radial oscillations, a precise peak remains stable over time which can hardly be explained by stochastically excited oscillations.

In the GCVS4, Semi--Regular variables are classified SRa (\emph{``Semiregular late-type giants with persistent periodicity and small light amplitudes''}), SRb (\emph{``Semiregular late-type giants with poorly defined periodicity or with alternating intervals of periodic and slow irregular changes, and even with light constancy intervals''}), SRc (\emph{``Semiregular supergiants''}) or SRd (\emph{``Semiregular variables of F, G, or K spectral types''}). We focus on SRa and SRb types. Table~\ref{tabfit1} and table~\ref{tabfit2} contain, among Miras, respectively 8 and 9 objects classified SRa or SRb in the GCVS4 five of them are SRb: U Boo, RY Cam, TV And, R Scl, Z UMa. Table~\ref{tabfit3} gives the result of satisfactory sine curve fitting for 28 additional SRb objects. Many Semi--Regulars even classified SRb show the presence of a clearly defined period throughout the entire observation period. In some cases, a more or less spread spectrum is also present.

The presence of a clear and persistent periodicity observed on the Miras, is also seen on many Semi--regulars although with a lower amplitude. The part of the spectrum allocated to stochastically excited oscillations has several larger peaks and a different central period. The non-radial and radial oscillations observed in Red Giants and some Semi--Regulars are similar in nature. The Semi--Regulars show both the solar--like oscillations of the red giants and Mira-type variations with lower amplitudes. There is enough evidence to suggest that Mira-type variations and solar--like oscillations are different in nature and thus Mira-type variations could hardly be attributed to pulsations but more likely to the presence of a companion.

\begin{figure}
  \centering
  \includegraphics[width=11cm]{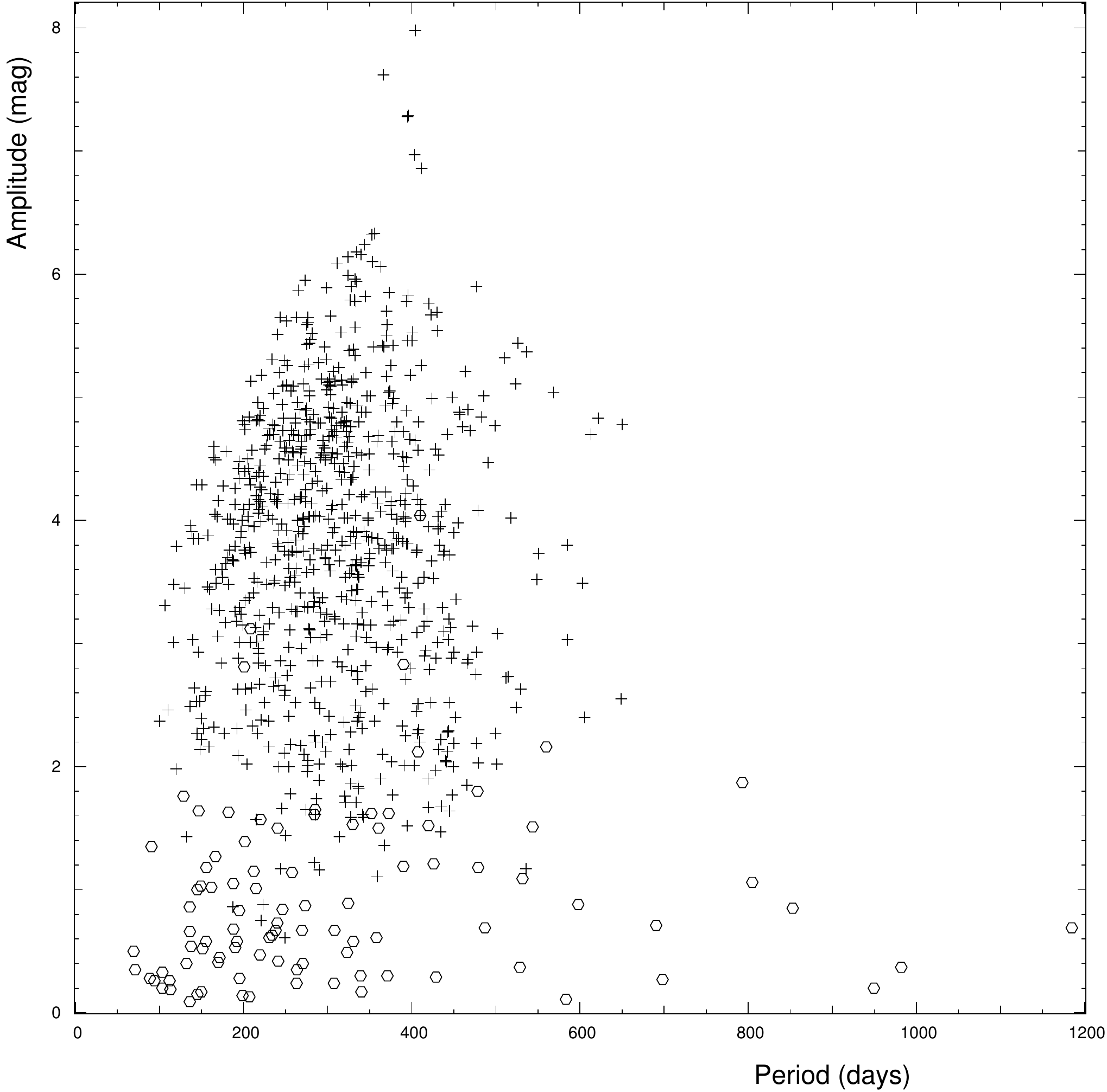}
  \caption{Amplitude versus period of Miras (plusses) and Semi--Regulars (circles) with sufficient data from the AFOEV and AAVSO databases. Periods and amplitudes are the results of a sine curve fitting on the whole dataset of each object.}
  \label{FigAmplvsPer}
\end{figure}

\subsection {Amplitudes}
Fitting sine curves on the light curves with sufficient data allows to draw on figure~\ref{FigAmplvsPer} the relationship between amplitudes and periods. In this figure, the clustering of Miras and Semi--Regulars simply reflects the definition of these objects. There is no gap between Miras and Semi--Regulars, as regards their main period, lower amplitudes than Miras. So Miras and Semi--Regulars are only distinguished by their definition on the diagram of amplitudes versus periods of their light curves and not by different properties.

Figure~\ref{FigAmplvsPer} shows a wide dispersion of amplitudes for a given period. 

In the frame of the pulsation paradigm, amplitudes are dependent on the driving and the damping of the oscillations, so we can expect similar amplitudes for objects with similar characteristics. It is not what is observed on figure~\ref{FigAmplvsPer}. 

The presence of a companion provides a range of amplitudes due to the random inclinations on the sky in accordance with what is observed (the simulation of eclipsing variables light curves will be discussed later in section 4.). 

\subsection{Some periods are varying with time}

\begin{figure}
  \centering
  \includegraphics[width=11cm]{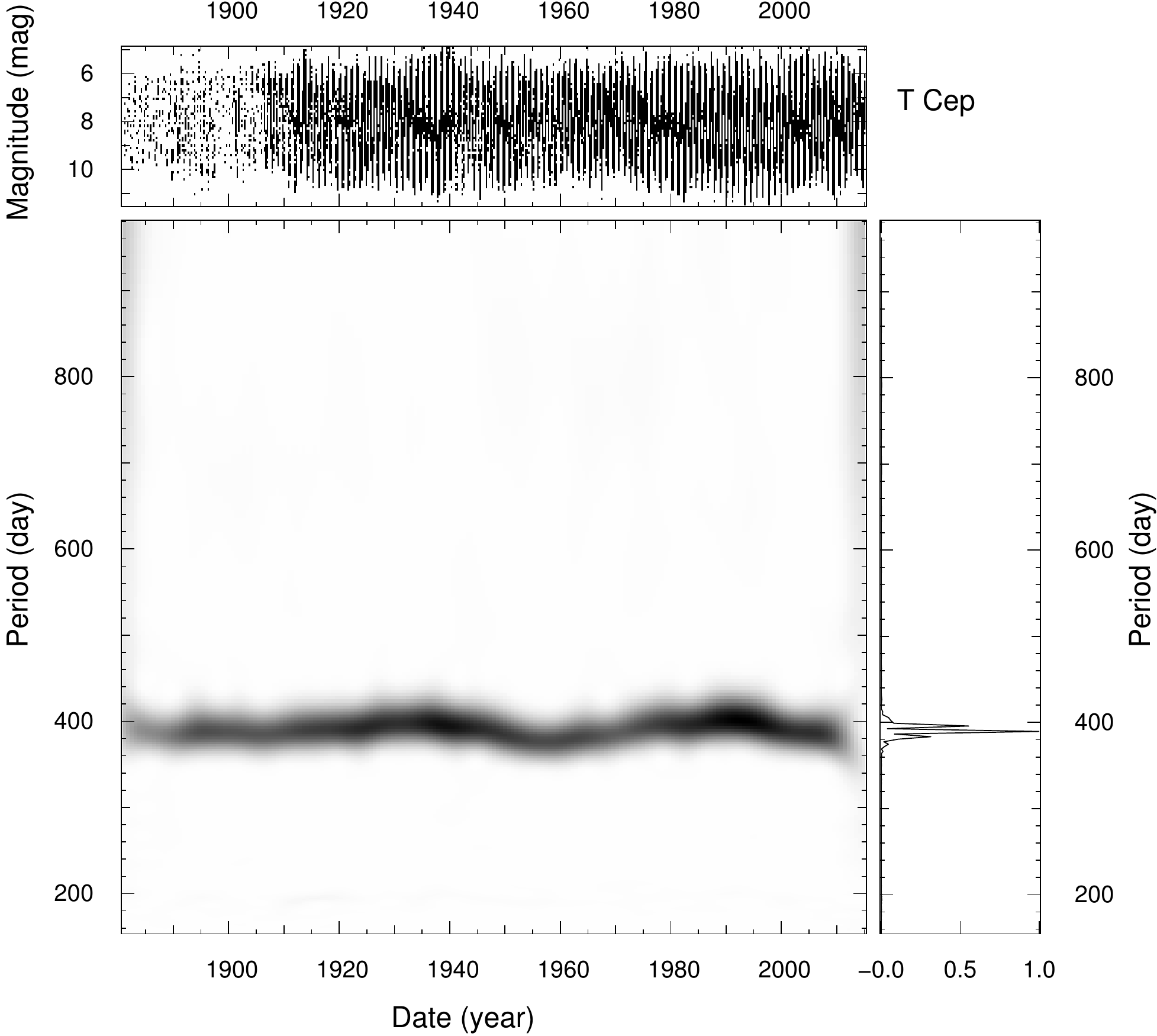}
  \caption{Time Analysis of T Cep light curve. Top panel: light curve from amateur databases. Central panel: time--period map. Right panel: power spectrum on the whole date set (scaled to maximum): period is varying between about 370 and 400 days along a cycle of about 50 years.}
  \label{FigTCep}
\end{figure}

\begin{figure}
  \centering
  \includegraphics[width=11cm]{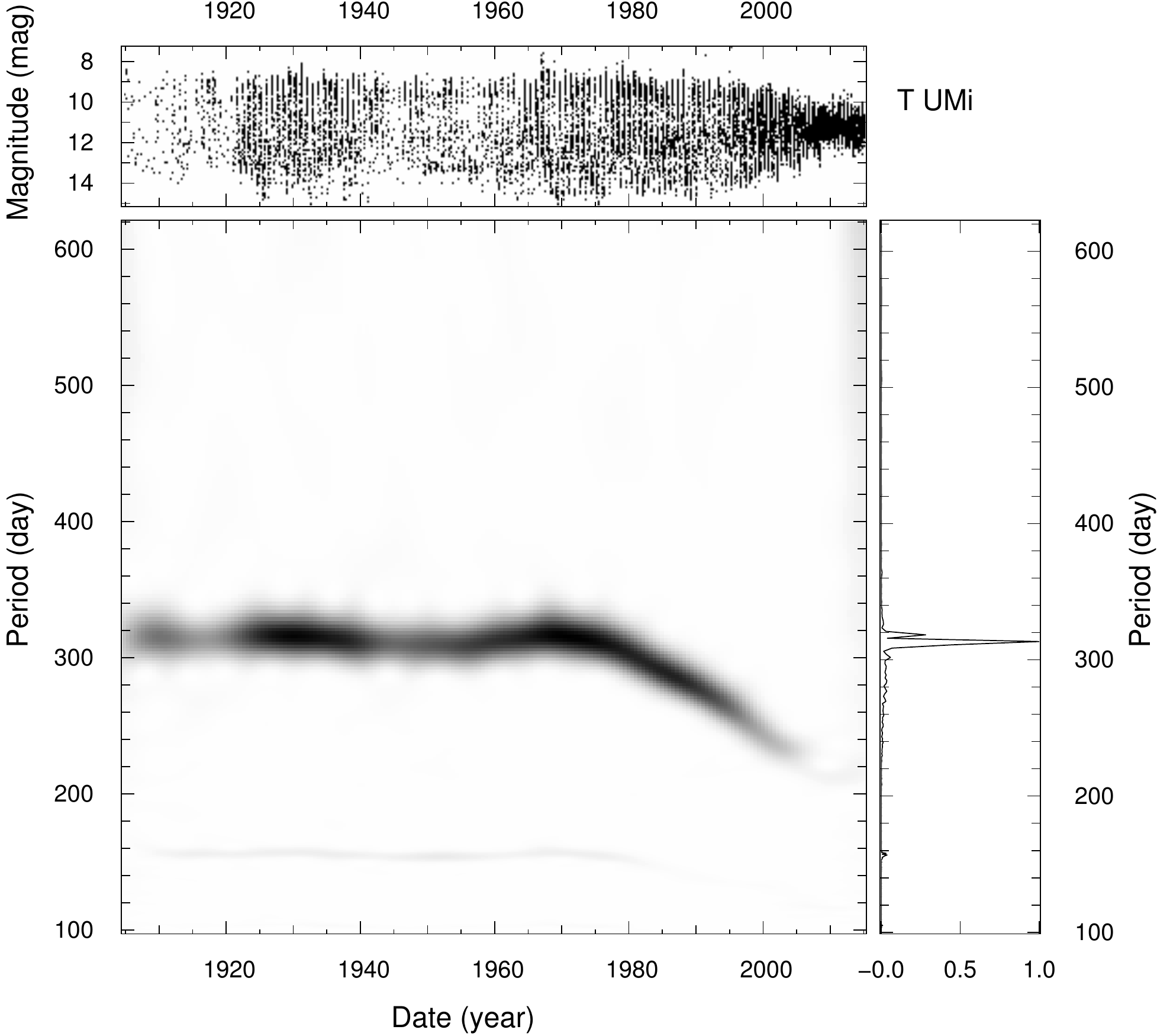}
  \caption{Time Analysis of T UMi light curve. Top panel: light curve from amateur databases. Central panel: time--period map. Right panel: power spectrum on the whole date set (scaled to maximum): strong period decrease from 1975, amplitude decrease from 2000.}
  \label{FigTUMi}
\end{figure}

The periods of some Miras do vary over time, such is the well-known case of R Aql whose period diminution rate is very stable: 0.47 days a year (Fig.~\ref{FigRAqlO-C} Top). The period is clearly defined even if it decreases over time. 
Some Miras show repetitive period variations. Fig.~\ref{FigTCep} gives an example of such variations. 

Some rare Miras show sudden variations. An extreme case is that of T UMi \citep{gal,szatmary} which shows a huge period diminution since 1976 (Fig.~\ref{FigTUMi}). From the year 2000 its amplitude has drastically decreased to the point that this star is likely not to present any more magnitude variations within a few years.

The pulsation paradigm explains the period variations as reflecting an underlying change in the star structure. The thermal perturbations induced by helium layer instabilities (helium shell flashes) change the physical conditions inside the star envelope and, in turn, the star period. \citet{woodzarro} tried some observed variations periods to make them fitting the star evolution models. The Miras whose periods are decreasing are supposed to be in the phase following an helium shell flash.

Taking binarity as the frame for interpreting, period variations are explained by angular momentum transfer inside the binary system through an exchange of matter between the two components, or outside the system through a mass loss. So a mass transfer from the more massive to the less massive star reduces the period, conversely transfer from the less to the more massive leads to a period increase. We may give an estimation of the mass transfer rate knowing the period variation rate. Applying the third Kepler law and assuming the angular momentum conservation of the system, we obtain:
\begin{displaymath}
\dot M_1 = - \dot M_2 = \frac{M_1 M_2}{3P (M_1 - M_2)} \dot P
\end{displaymath}
where $M_1$ and $M_2$ are respectively the masses of the more and the less massive components, $\dot M_1$ and $\dot M_2$ the rates of mass variation, $P$ the period of the system, and $\dot P$ its variation with time. 

We find a period variation of $0.48$ day by year for R Aql and a $274$ days period. With a mass of $1M_{\sun}$ for the massive component, the mass transfer rate is about $1.75\, 10^{-3}M_{\sun}$ by year for a mass ratio of $0.5$ and $3\, 10^{-5}M_{\sun}$ by year for a mass ratio of $0.05$. It should be noted that R Aql is an extreme case, very few Miras show a so strong period decrease.

The Miras with their exceedingly extended atmosphere generate a mass transfer to the companion when it gets into the atmosphere leading to a period decrease. We may think it is the case for T UMi and we can then witness the fall of the companion into the red giant atmosphere leading this companion to disappear.

\subsection{Double humps}

\begin{figure}
  \centering
  \includegraphics[width=10cm]{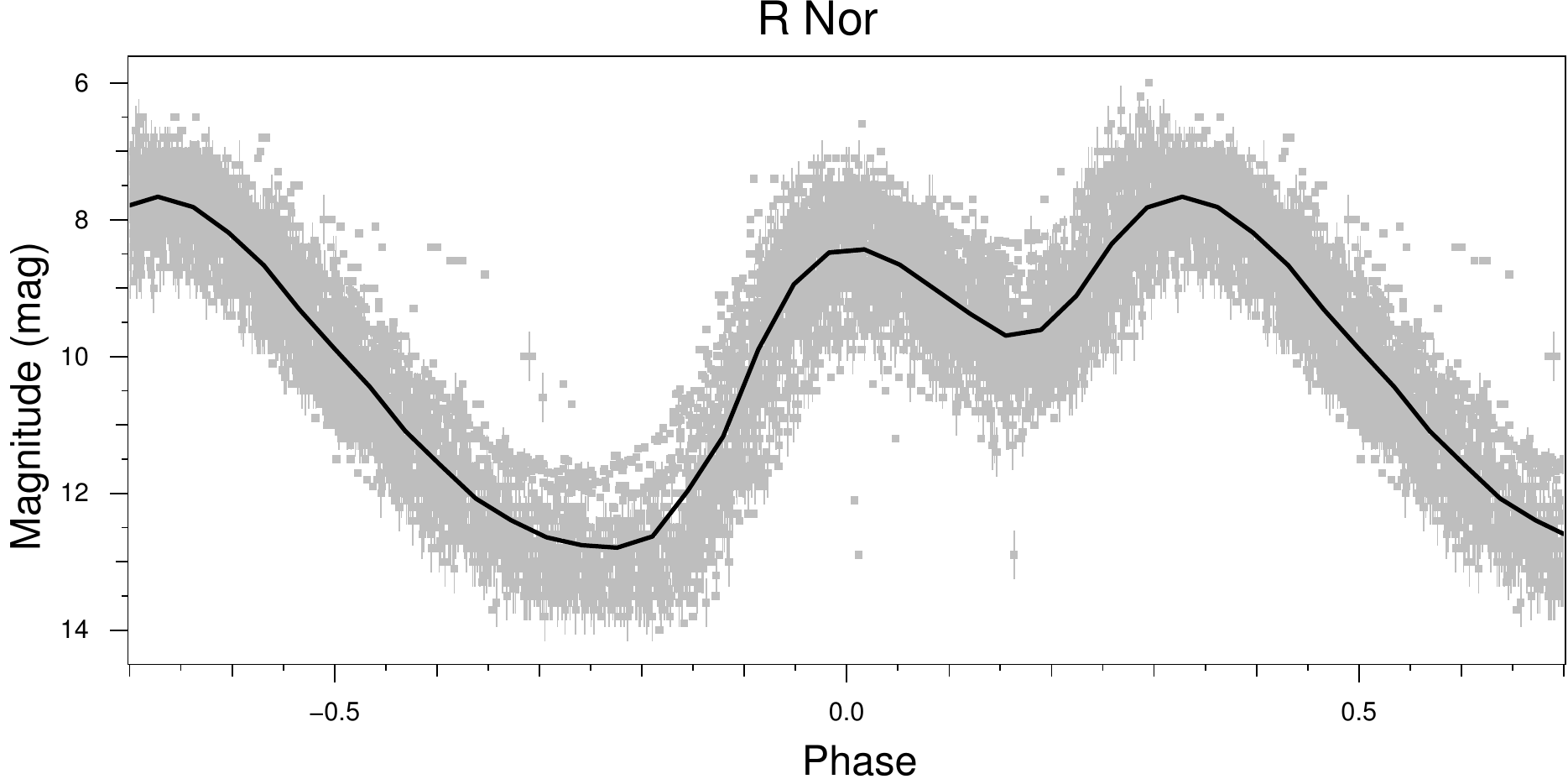}
  \includegraphics[width=10cm]{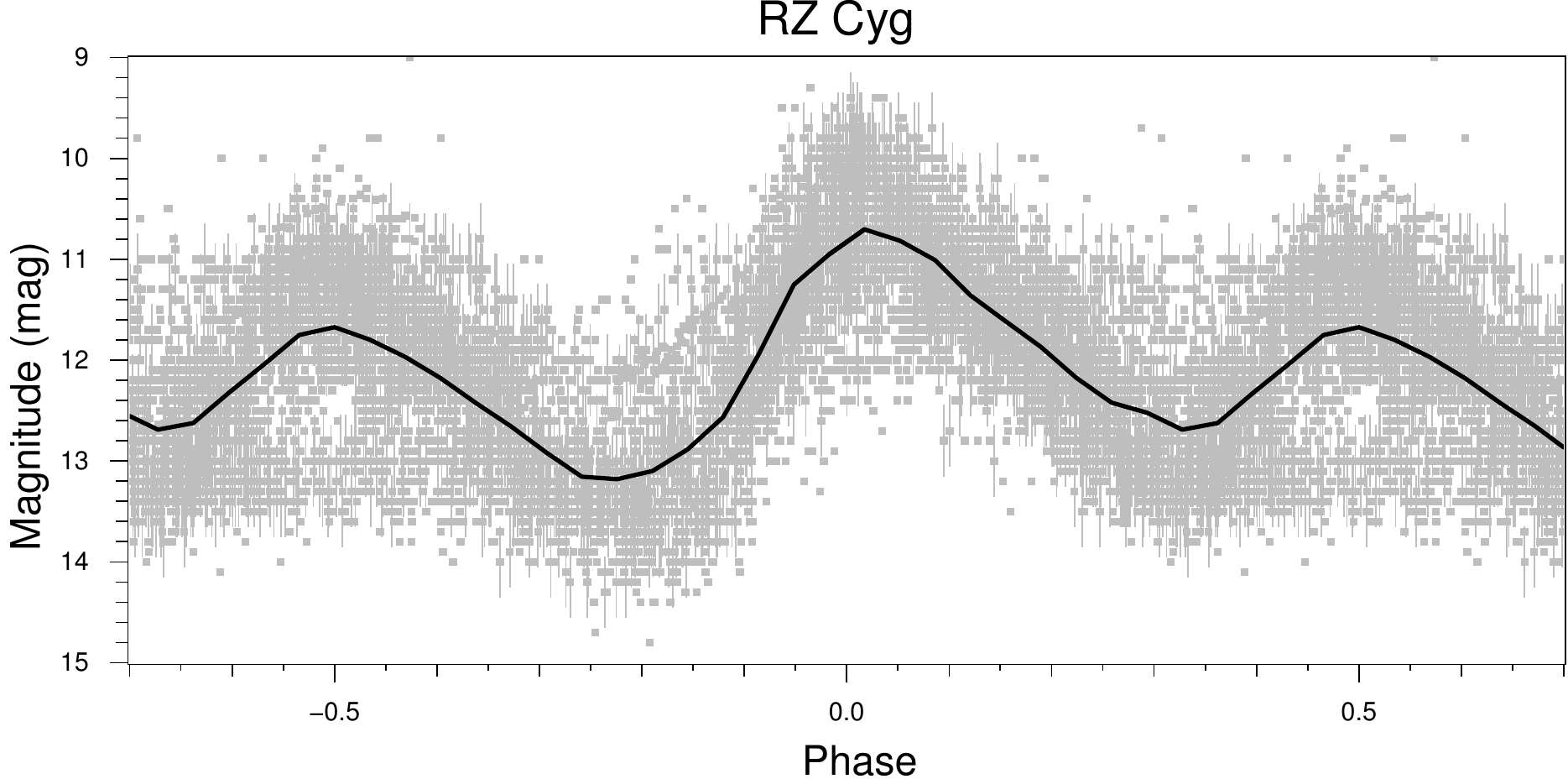}
  \includegraphics[width=10cm]{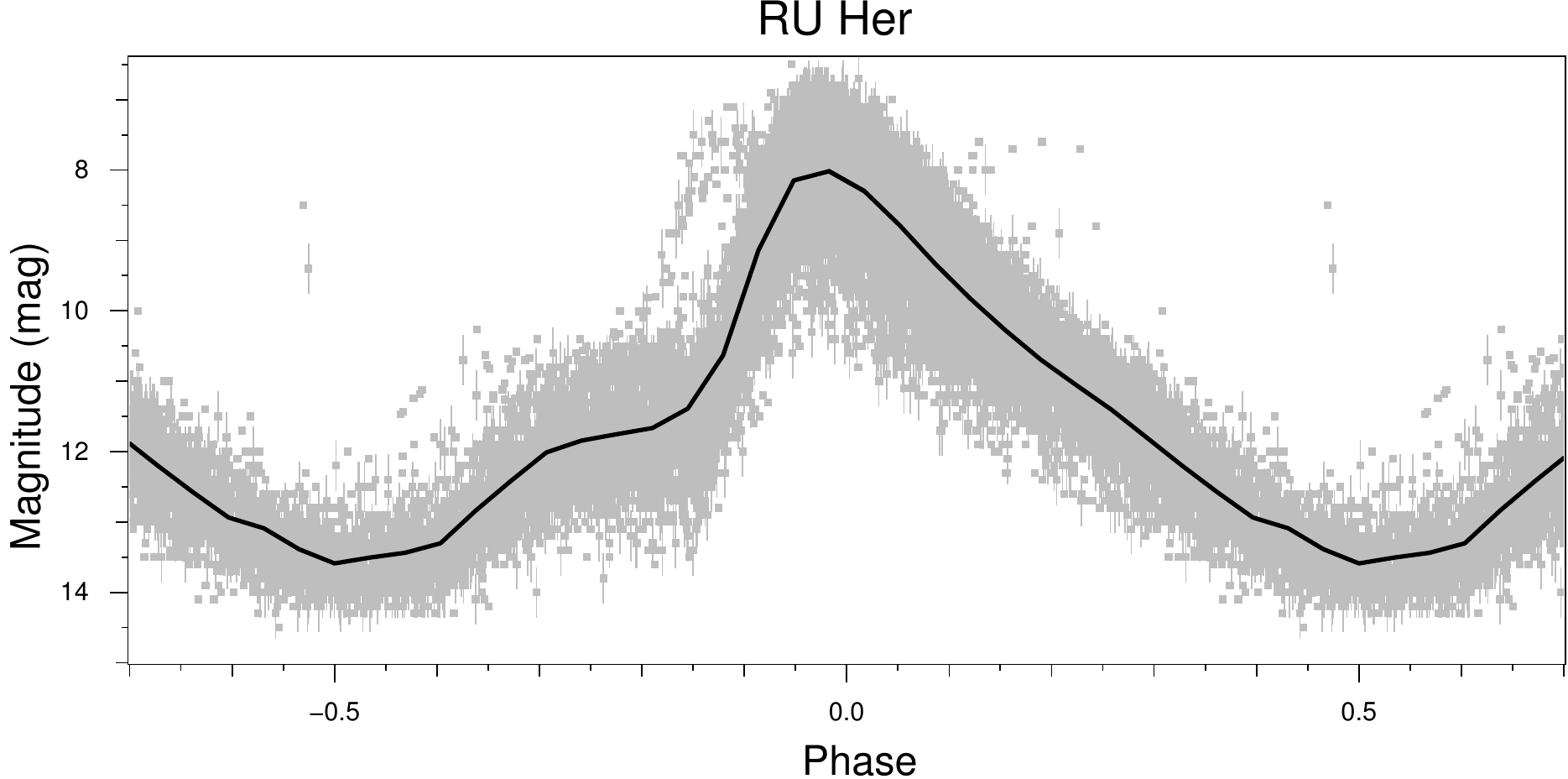}
  \includegraphics[width=10cm]{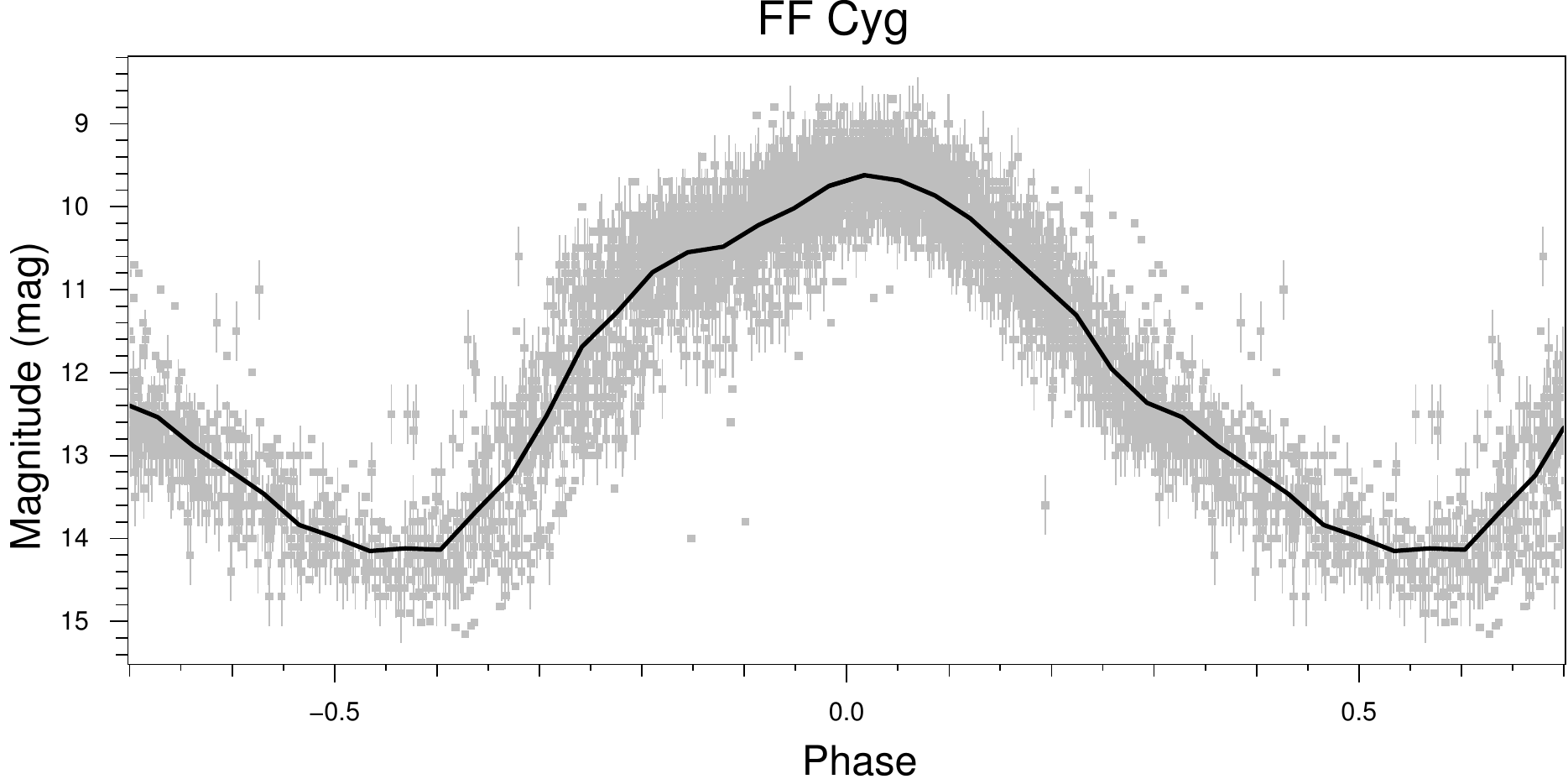}
  \caption{Folded visual light curves of Mira variables showing double humps or bumps. Top left. R Nor period 518 days. Top right. RZ Cyg period 543 days. Bottom left. RU Herculis period 485 days. Bottom right. FF Cyg period 325 days. Observed magnitudes (Gray marks) were folded according to the periods. Mean folded light curve (black line)}
  \label{Fig2bosses}
\end{figure}

Many Miras light curves may be fitted to a sine curve, or rather to two half sine curves, the rising time from minimum to maximum being usually shorter than from maximum to minimum. Some rather rare objects show a light curve with two maximum. It is the case of R Cen, R Nor or RZ Cyg, all these variables have rather long periods of about 500 days. Fig.~\ref{Fig2bosses} top shows two examples of visible light curves with two humps by period. 

Other more numerous Miras show a ``bump'' on the rising from the minimum to the maximum light. One observes a continuum from sinusoidal light curves to double humps light curves through curves showing more and more obvious bumps. Fig.~\ref{Fig2bosses} bottom: two examples of visible light curves showing bumps. 

To explain these double-humped light curves using the pulsation paradigm, \citet{lebzelter2005} suppose two radial pulsation modes are simultaneously excited: fundamental and first harmonic or two harmonic modes. These two modes having necessarily periods of an exact 2:1 ratio: \emph{``the bumps are the result of a 2-to-1 resonance between the fundamental mode and the first overtone''} \citep{lebzelter2005}.

Double humps are easy to interpret in the frame of binarity, the light curves showing two successive minimum which are related to the eclipses of both components. Some examples of synthetic light curves are given below.

\citet{wood2000} using the $K$ photometry of variables in the LMC from the MACHO database finds five sequences in the $K$--log period diagram. The sequence C is attributed to Miras and the sequence E, on the grounds of their double humps, to close eclipsing or contact binaries. \citet{soszynski2011} find the same sequences for the long period variables of the Small Magellanic Cloud from the OGLE-III Catalogue of Variable Stars and identify the sequence E objects as close binaries with ellipsoidal shapes; the periods of these stars correspond to half of the orbital periods. Therefore when the sequence E is plotted with periods multiplied by 2 -- the orbital periods -- it appears to be the same as sequence D and so sequence D and E objects would be binaries \citep{soszynski2007}. 
\citet{wood2010} supposes that these sequence E objects, when evolving to larger dimensions, fill their Roche lobes and could be the precursors of planetary nebulae. However \citet{wood2010} challenges the idea that the sequence D objects would be binaries on grounds that the light and velocity curves of sequence E and sequence D are different : the amplitudes of the sequence D velocity curves is around $3.5\,\mathrm{km.s^{-1}}$ an order of magnitude smaller than sequence E, and there is one minimum of the light curve for each cycle of the velocity curve in sequence D stars, on the contrary E variables show two minimum of the light curve for each cycle of their velocity curve. \citet{soszynski2011} show that O-rich and C-rich Miras and Semi--Regulars populate not only the sequence C but also the sequence D of the LMC (their figure~5). We conclude that the sequence E overlaps the sequence D and that Miras, Semi--Regulars and close binaries share these sequences.

\subsection{Infrared light curves}
\begin{figure}
  \centering
  \includegraphics[width=11cm]{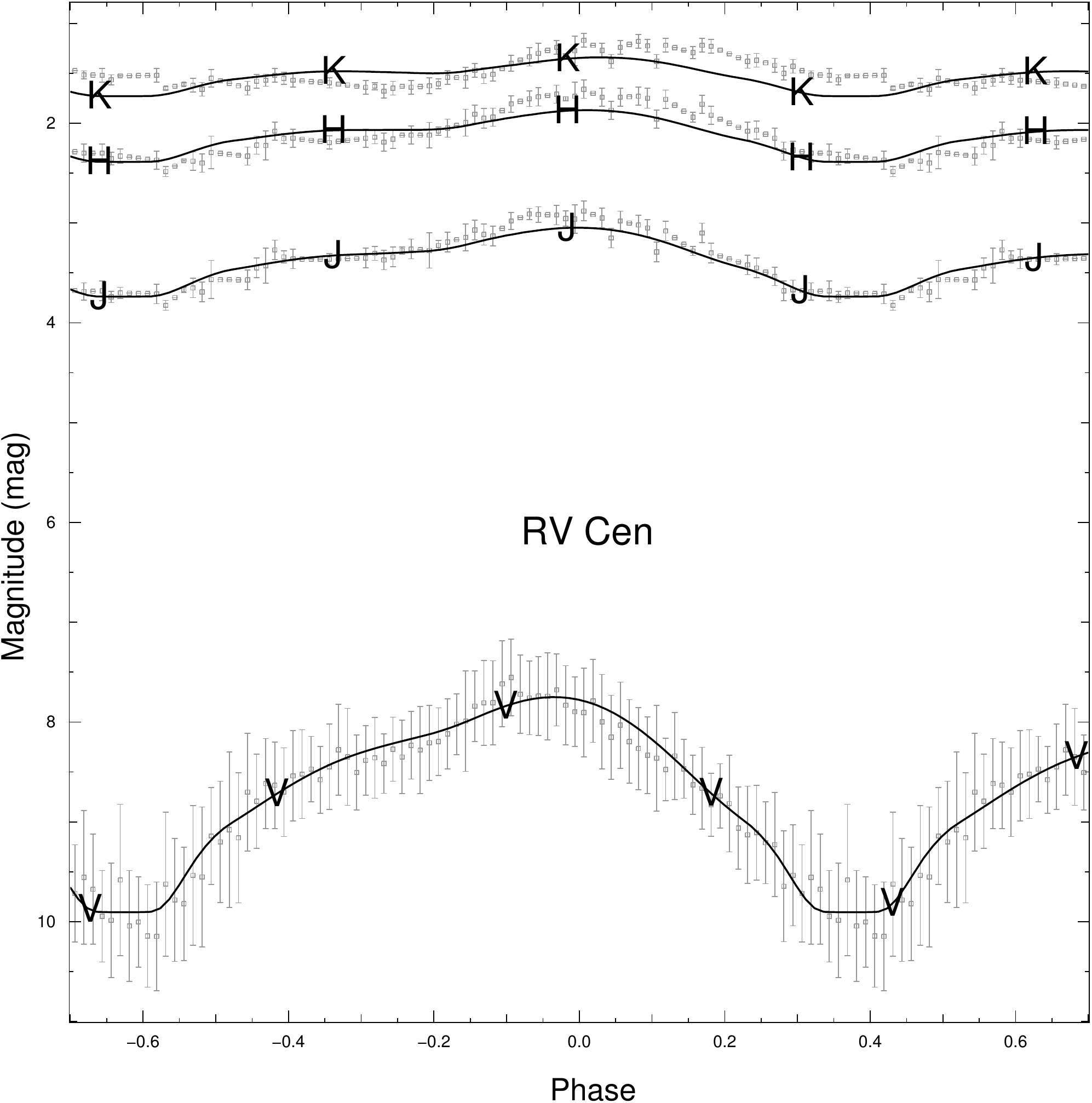}
  \caption{RV Cen visible and infrared light curve. 'V' Visible light curve, 'J', 'H' and 'K' near infrared light curves. Dot : observed folded mean magnitudes from AAVSO and AFOEV (visual) and \citet{whitelock2000,whitelock2006} (infrared). Solid lines are the result of fitting the mean observed visible light curve with the help of NightFall software. Parameters are given table~\ref{tabfitRV}.}
  \label{FigRVCen}
\end{figure}

Double humps on the visible light curves can also be seen on infrared light curves. Moreover, one sees some infrared light curves with double humps although visible light curves show no double humps. An example is given Fig.~\ref{FigRVCen}. 

The pulsation paradigm gives no explanation of these differences between visible and infrared light curves.
 
Synthetic light curves of eclipsing binary stars can be computed by software based upon the \citet{wilson} code. The Mira RV Cen shows a bump in the visible; its light curve is reproduced on Fig.~\ref{FigRVCen} with the help of the NightFall software\footnote{http://www.hs.uni-hamburg.de/DE/Ins/Per/Wichmann/Nightfall.html}. Table~\ref{tabfitRV} gives the fitting parameters. The resulting computed light curves in the infrared for the $J$, $H$ and $K$ bands show amplitudes and shapes close to the observed ones. The presence of a hotspot on the companion's surface near the Lagrangian point L2 is assumed; Long Period Variables are losing mass, therefore it makes sense that mass overflow through the Lagrangian point L2 may produce such hotspot (see discussion below).

\begin{table}
\caption{Fitting parameters of the RV Cen synthetic light curve}
\label{tabfitRV}
\begin{tabular}{@{}l c}
\\
  \hline
  \hline
  Type: Contact binary \\
  Mass ratio & 0.04 \\  
  Incidence & $ 63\degr$ \\
  Primary temperature & 2046 K\\ 
  Secondary temperature & 2594 K \\
  \hline                        
  Spot secondary: \\      
  Longitude & $8\degr$ \\
  Radius & $70.6\degr$ \\
  Dimfactor & $1.19$ \\
  \hline
  Spot primary: \\ 
  Longitude & $25\degr$ \\
  Radius & $38.4\degr$ \\
  Dimfactor & $1.19$ \\
  \hline
  \hline
\end{tabular}
\end{table}

\subsection{Shapes are changing}
The light curves are not identical during successive cycles. The shape of the light curve may change while the minimum and maximum remain relatively constant, or the shape of the curve remains about the same but its level changes (obscuration events). 

\begin{figure}
  \centering
  \includegraphics[width=11cm]{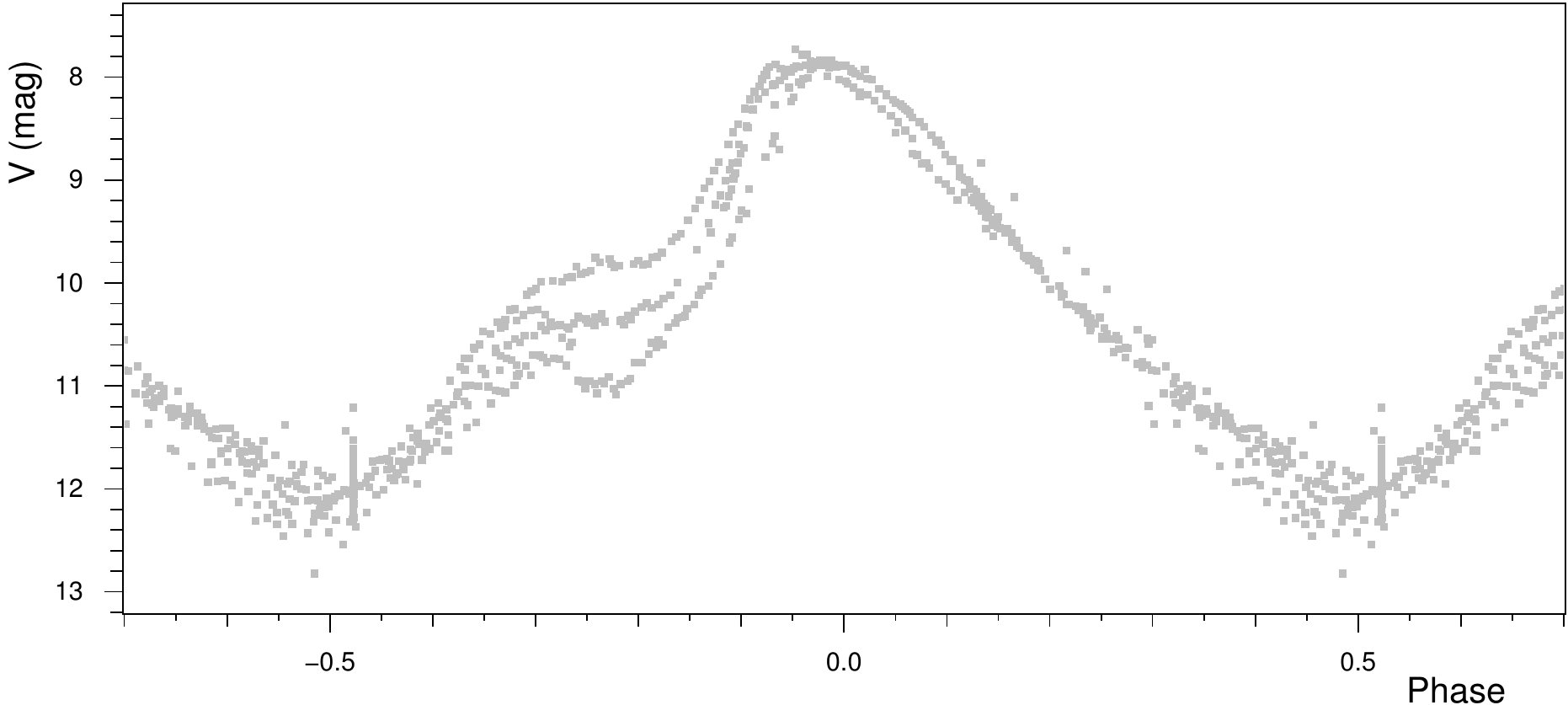}
  \caption{Folded light curve of S Lep. The ASAS data more precise than the amateur data show the variation of the shape of the light curve of S Lep from one cycle to another. The main variation concerns the ``bump''.}
  \label{FigS_Lep}
\end{figure}

\begin{figure}
  \centering
  \includegraphics[width=11cm]{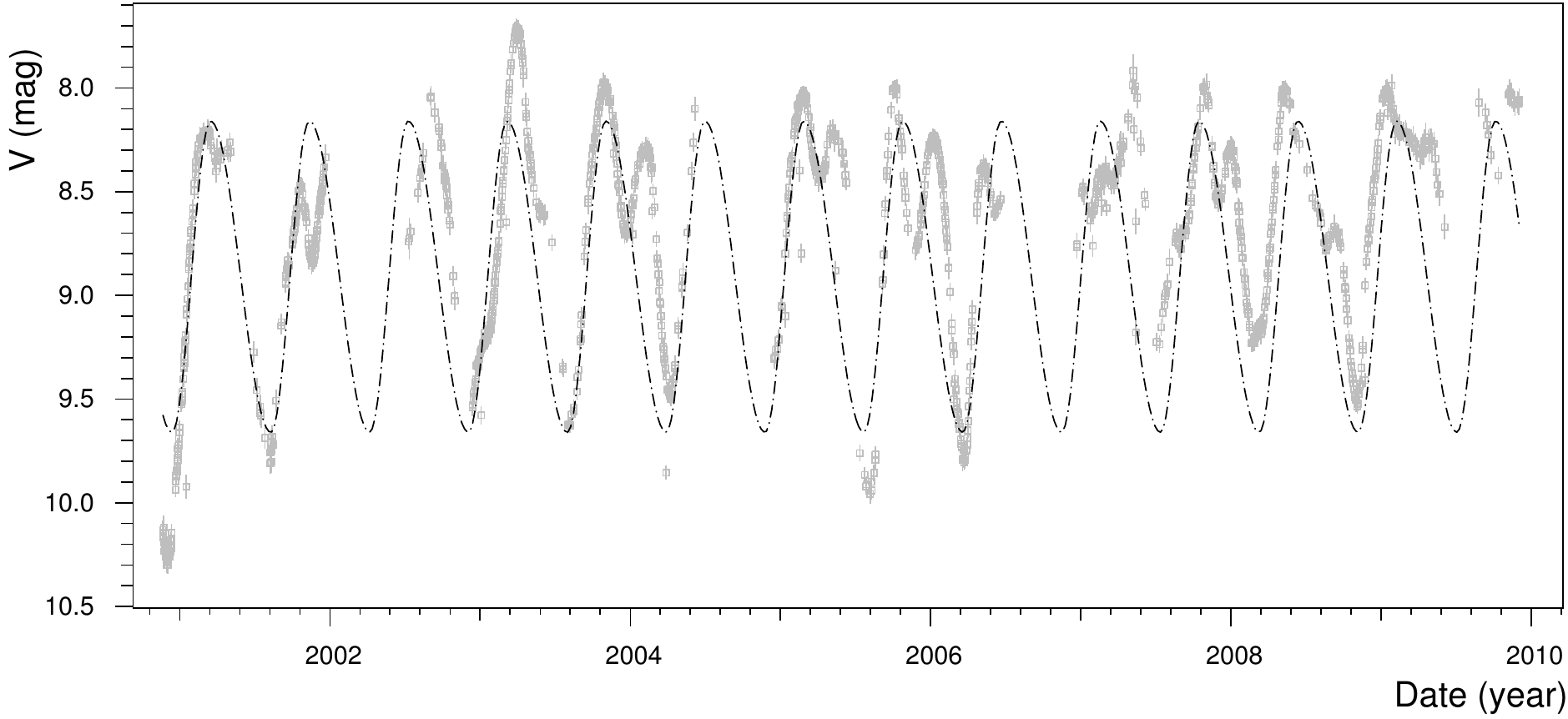}
  \caption{Light curve of R Men drawn with the ASAS data (Gray). The black dash dot line is the sine curve fitted on the whole dataset. From cycle to cycle the shape of the light curve is changing.}
  \label{FigR_Men}
\end{figure}

\subsubsection{Shapes.}
Fig.~\ref{FigS_Lep} shows the variation of the shape of the S Cep light curve which depends on the cycle. With the precision of the ASAS data the bump appears more or less prominent. Another example is the light curve of R Men (Fig.~\ref{FigR_Men}) showing very different shapes along different cycles: double humps, smooth curve, or bump. These variations have no true understanding in the frame of the pulsation paradigm. We interpret them as variations of a hotspot on the secondary.

\subsubsection{Obscuration events.}
\begin{figure}
  \centering
  \includegraphics[width=11cm]{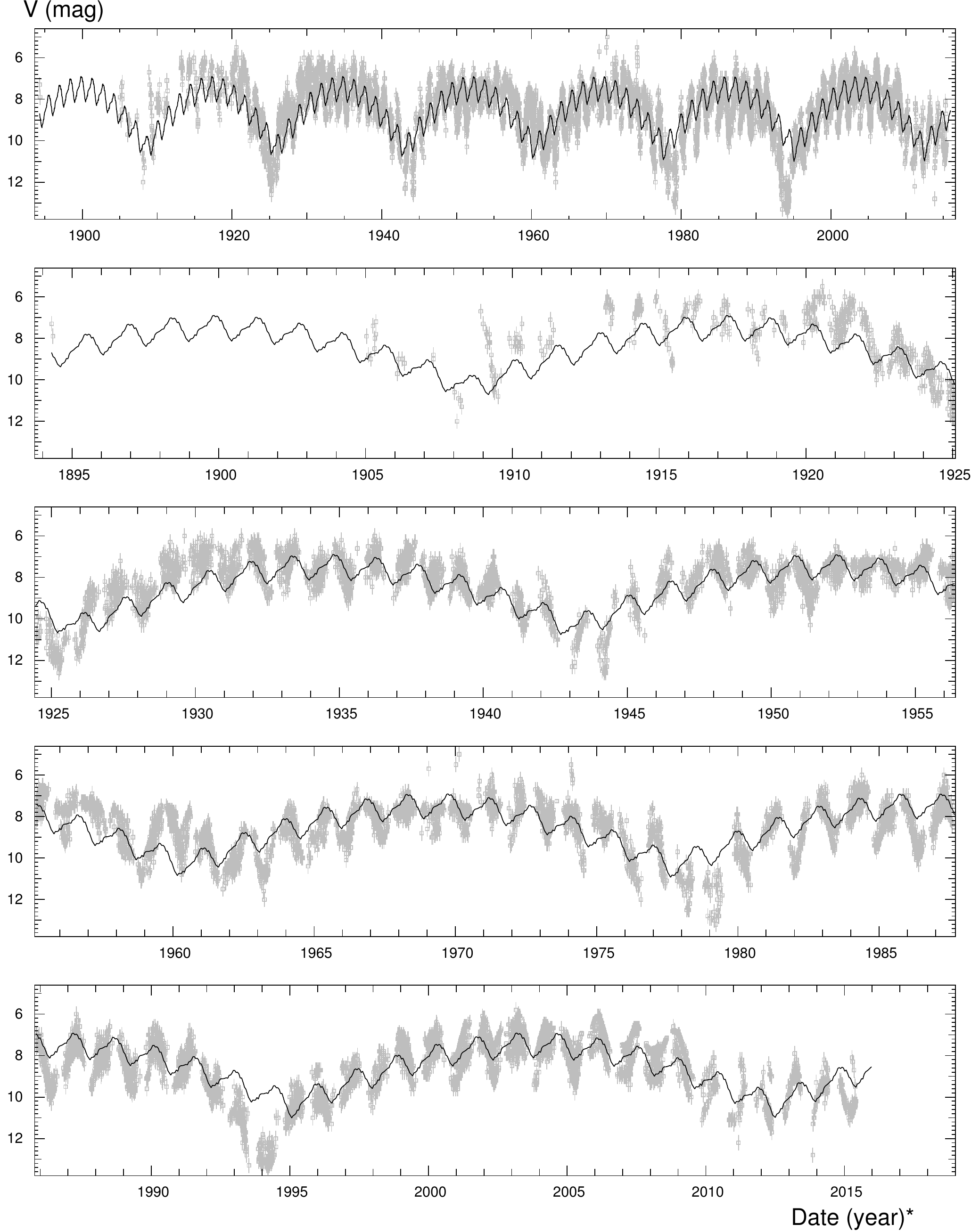}
  \caption{Visual light curve (Gray) of V Hya showing regular obscuration events. The black line is the result of a fit whose parameters are given in given table~\ref{tabfitVHya}. Top panel: full data set, Following panels: successive zooms.}
  \label{FigV_Hya}
\end{figure}

Some light curves appear to be more or less periodically dimmed. Most striking is the V Hya light curve (Fig.~\ref{FigV_Hya}) showing a variation with a period of about 17 years which is overlaid to a ``normal'' variation of 531 days. We fitted this light curve with the parameters given in Table~\ref{tabfitVHya}. The long-term variation is interpreted as an obscuration by matter orbiting the Mira.

\begin{figure}
  \centering
  \includegraphics[width=11cm]{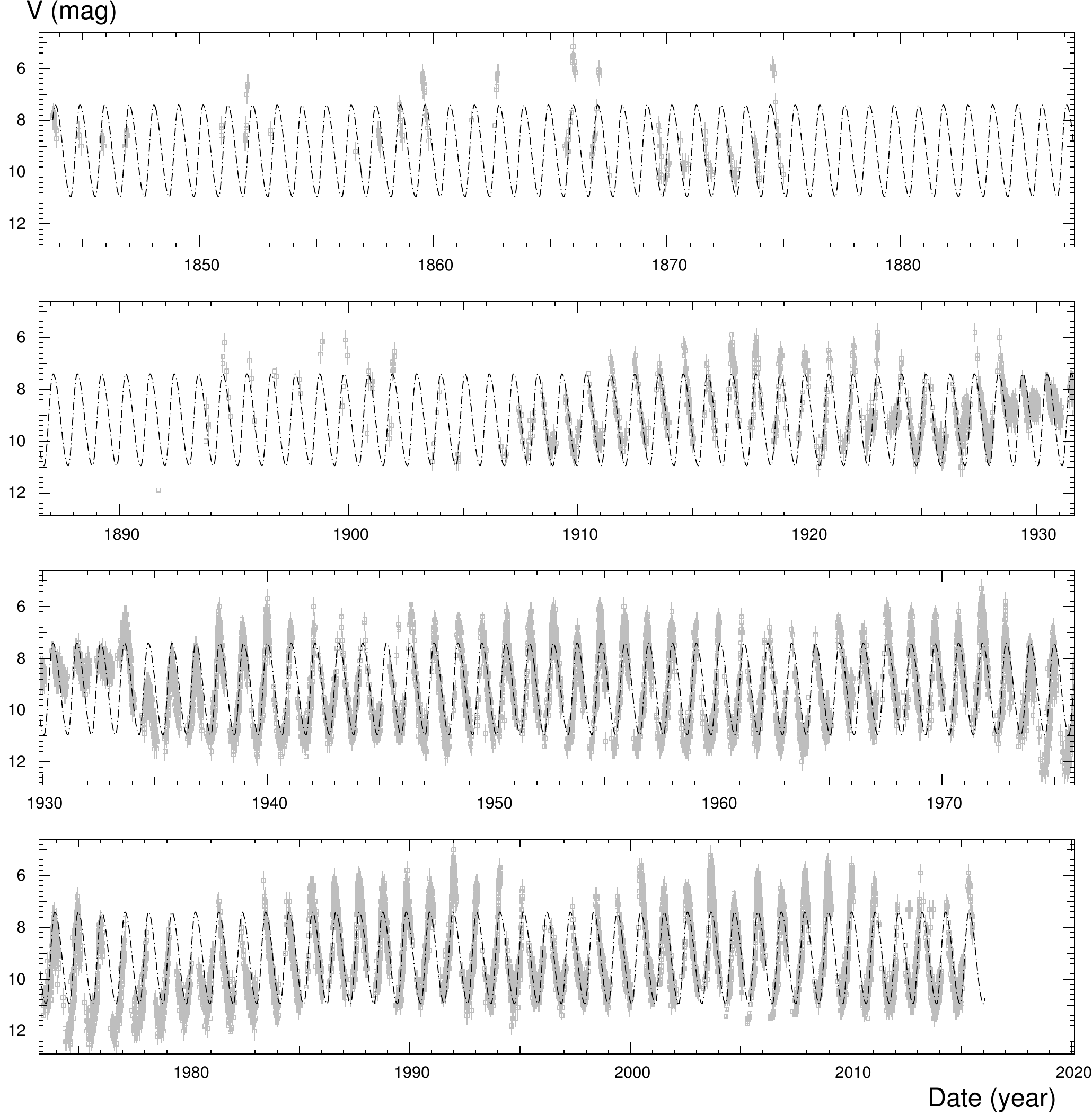}
  \caption{Visual light curve (Gray) of R Aqr showing irregular variations. The dash dot line is the result of a constant parameters fitting.}
  \label{FigR_Aqr}
\end{figure}

Some other Miras show irregular variations. Fig.~\ref{FigR_Aqr} shows the light curve of the symbiotic Mira R Aqr; it is hard to distinguish between the intrinsic variations of the Mira and external obscurations.

To conclude this section, we try to show that the pulsation paradigm gives certainly not a better explanation of the features of the light curves that the companion hypothesis.

\begin{table}
\caption{Fitting parameters of the V Hya synthetic light curve}
\label{tabfitVHya}
\begin{tabular}{l c}
\\
  \hline
  \hline
  \noalign{\smallskip}
  Long term variation: & half sine wave \\
  Period (day) & 6321. \\ 
  Date of maximum (HJD) & 2456081 \\ 
  Amplitude (mag) & 2.90\\
  \hline
  Short term variation: & 2 half sine waves\\
  Maximum (mag) & 7.0 \\ 
  Minimum (mag) & 8.1 \\ 
  Date of maximum (HJD) &  2451692\\ 
  Asymmetry factor & 0.69 \\
  Period (day) &  531. \\ 
  Period variation (day/year) &  -0.015 \\ 
  \hline
  \hline
\end{tabular}
\end{table}

\section{Modelling the light curve}

\subsection{Pulsation}
The first attempts to model pulsating stars were to study the instability conditions of radial modes (1-D models) \citep{eddington1918}.

Since then models have advanced with a major shift from the $\varepsilon$-mechanism to the $\kappa$-mechanism as excitation mode (see discussion below), then to non-adiabatic models, non-linear models, taking account for non-radial modes, and at last to 3-D models. At the same time, models with \emph{piston} boundary conditions were developed.

\subsubsection{Models of radial modes.}
Regarding Miras, \citet{wood2007} considers that the problem is that most of the heat transport in the envelope of the Miras is by convection where the pulsation is confined. Linear, non-adiabatic pulsation models allow to get an estimation of the pulsation period. Detailed models of \citet{xiong1998} show growing stability with decreasing effective temperature, by contrast observations are showing that the coolest objects with the longest periods have the largest amplitudes. \citet{wood2007} concludes that \emph{``current convection theories do not lead to reliable quantitative predictions regarding model stability''}. 
A more sophisticated approach is that of non-linear pulsation models. The problem with these models is that they lead to a too strong driving (i.e. a too large amplitude compared to what is observed).
To avoid the problem, a damping effect is introduced using a suitable adjustable parameter of turbulent viscosity whose value is adjusted to get the observed amplitudes \citet{wood2007}.
The periods obtained with these models can be, depending on the case, longer or shorter than those obtained by linear calculation. So, despite numerous efforts at improving the models, the results are not satisfactory for Miras.

\subsubsection{Models of radial and non-radial modes.}

Based on solar observations, the theory of stellar oscillations made many advances with the data from CoRoT and \emph{Kepler} missions which allow to validate these advances on objects other than the Sun \citep{tassoul}. 
The observed power spectra of red giant light curves are rich with many radial and non-radial modes excited to observable amplitudes. It can be expected that only the lowest levels of angular modes ($\ell<3$) are indeed observed: the higher degrees of angular modes are not visible because they are averaged when viewed on an entire hemisphere of the star (when the surface of the star is not resolved).

The most striking observation is the comb--like structure of the power spectrum of the red giant light curves  \citep[for instance, see the fig.~4 of][]{stello}.
These structures can be characterised by the frequency of the maximum and by the distances between lines. The predictions may be compared with these observed characteristics of the spectrum. 

There are two types of modes: p-- and g--modes. The restoring force of p--modes is pressure. They are located in the envelope of the star, have spacing between orders almost identical RV Cen synthetic in frequency space. The restoring force of the g--modes is gravity. They are located in the centre of the star, and have spaciness between orders almost identical in period space. Some g--modes in the centre of the star may couple to p--modes producing observable amplitudes. They are the so-called mixed modes which are important clues in determining the characteristics of the star. Mixed modes are only non-radial modes like g--modes.

The observation of a comb--like power spectrum allows linking theory and observation. However, the theory can predict neither the amplitude, nor the shape of the red giant light curve and even more so for Miras or Semi--Regulars. The amplitudes are the result of a balance between driving and damping mechanisms originating in the turbulence induced by convection \citet{chaplin} and are impossible to estimate.

Furthermore the theory does not explain how the spectra of red giants showing multiple radial and non-radial modes with low amplitudes could evolve into Miras spectra having a single radial mode with a very high amplitude.

\subsubsection{Models with driven pulsation.}
 Since the work of \citet{bowen}, numerous studies modelled the Miras with a \emph{piston} boundary condition. The piston is located outside the presumed excitation zone of the pulsation but in an optically thick layer. This approach is made necessary by the difficulty to achieve self--consistent models of pulsating Miras. This method overcomes the unsatisfactory modelling of the pulsation mechanism by forcing the pulsation with the \emph{piston} boundary conditions of the model.

These models allow a fairly good reproduction of molecular line profiles \citep{nowotny2010} and of the light curves \citep{nowotny2011}, at least of carbon Miras. The limitation is the free parameters of the driven pulsation: amplitude and period of the piston.

\subsubsection{Three-dimensional (3D) models.}
Radiative and hydrodynamic 3D models, first applied to the sun, are now applied to red giants and supergiants. In this case, the model can not be confined to a box in the star but must include the whole star volume ('star in a box') \citep{freytag2002,woodward,brun,chiavassa}.
The 3D simulations of red giants show a few huge convective cells on the surface of the star \citep{woodward,freytag2008,brun}. 

\citet{cruzalebes} observed departure from centrosymmetry of the photospheres of 16 bright late-type stars that they attribute to large convective cells which were predicted previously by \citet{schwarzschild}.

Convective velocities obtained are of several Mach number, and form shock waves at the surface of the AGB stars \citep{freytag2008,freytag2012}. The importance of these convection fluxes and the kinetic energy they carry are at odds with the traditional treatment of convection by the Mixing Length Theory (MLT) \citep{brun}.

3D model atmospheres of red giants, as they represent a detailed granulation pattern, begin to be able to reproduce the photometric variability, and the power spectra of this micro-variability could be soon confronted to observed ones \citep{ludwig}.

These simulations do not reproduce radial pulsations, but more or less periodic variations, in shape and volume, following the emergence of large convective cells \citep{freytag2008}. 

\begin{figure}
  \centering
  \includegraphics[width=11cm]{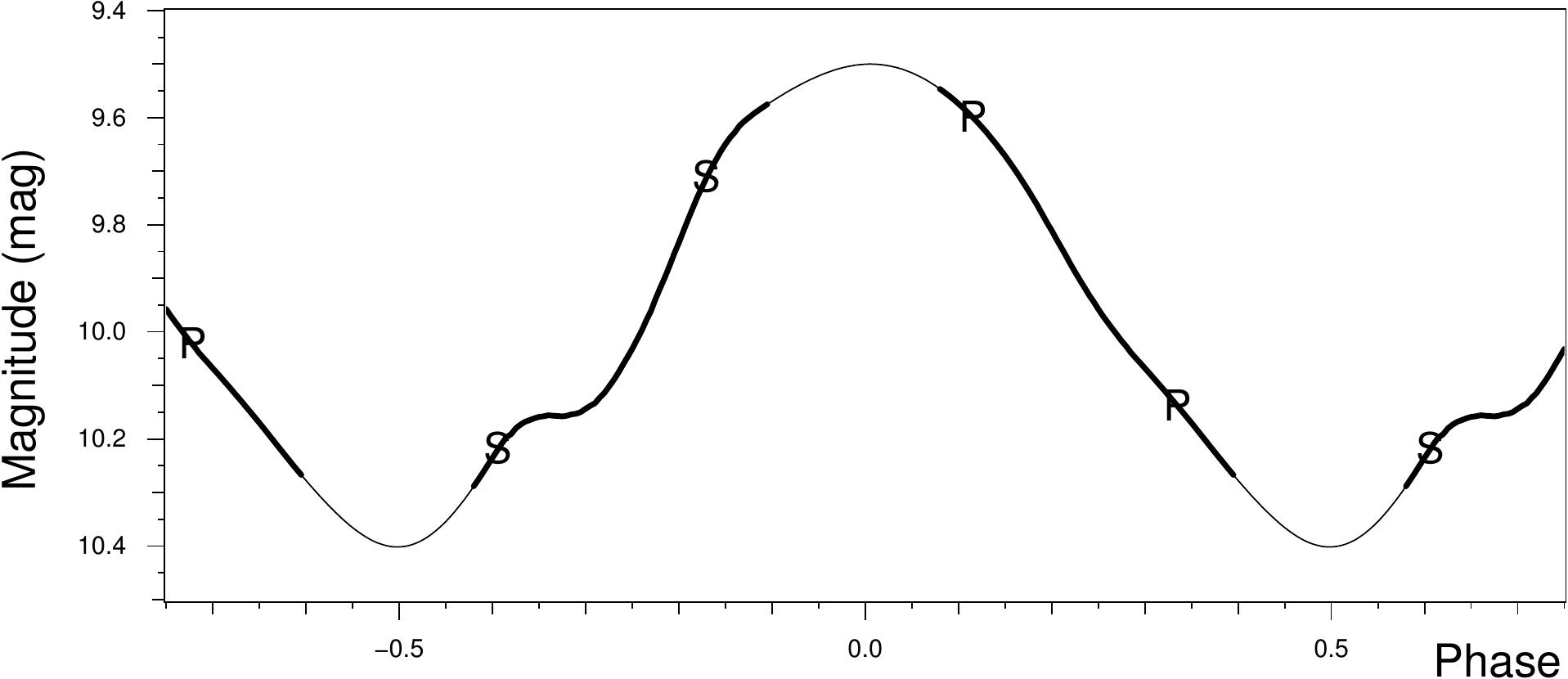} 
  \includegraphics[width=11cm]{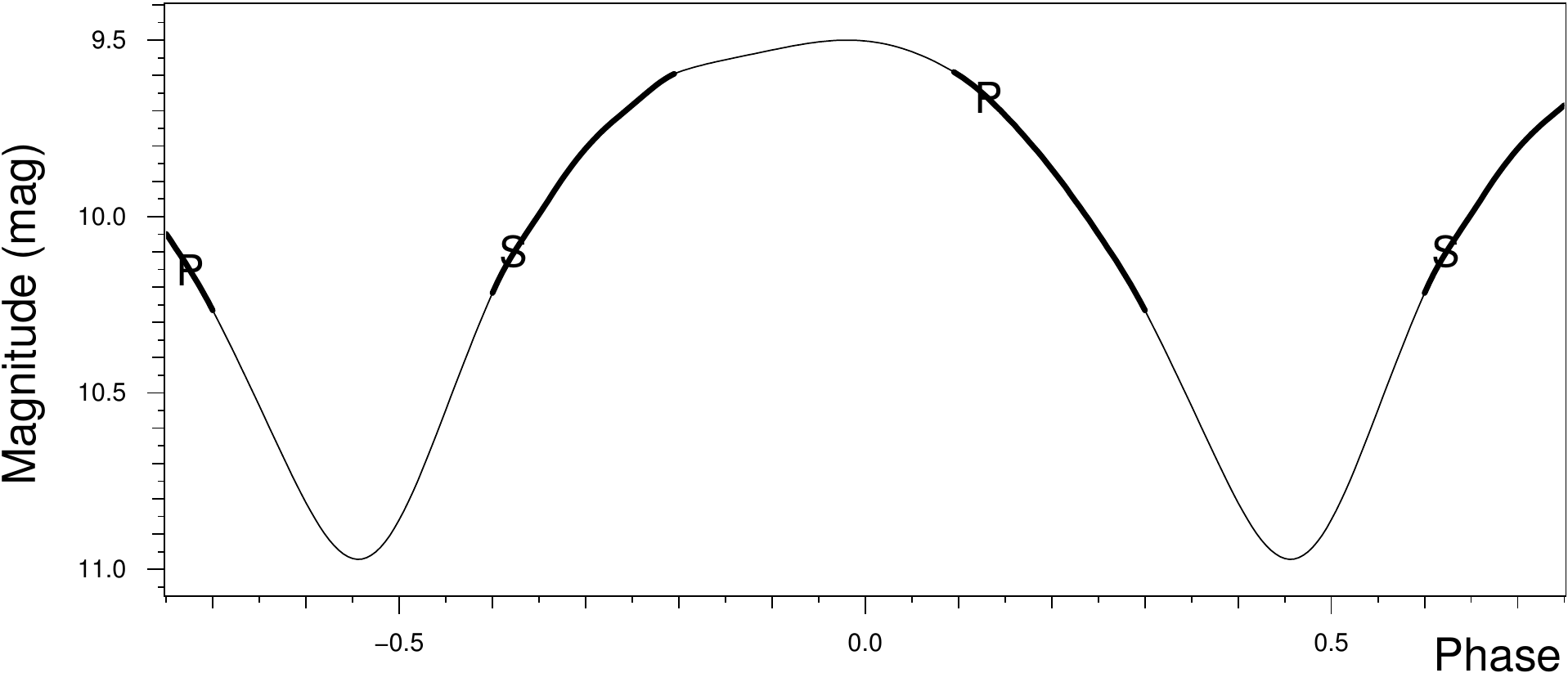}
  \includegraphics[width=11cm]{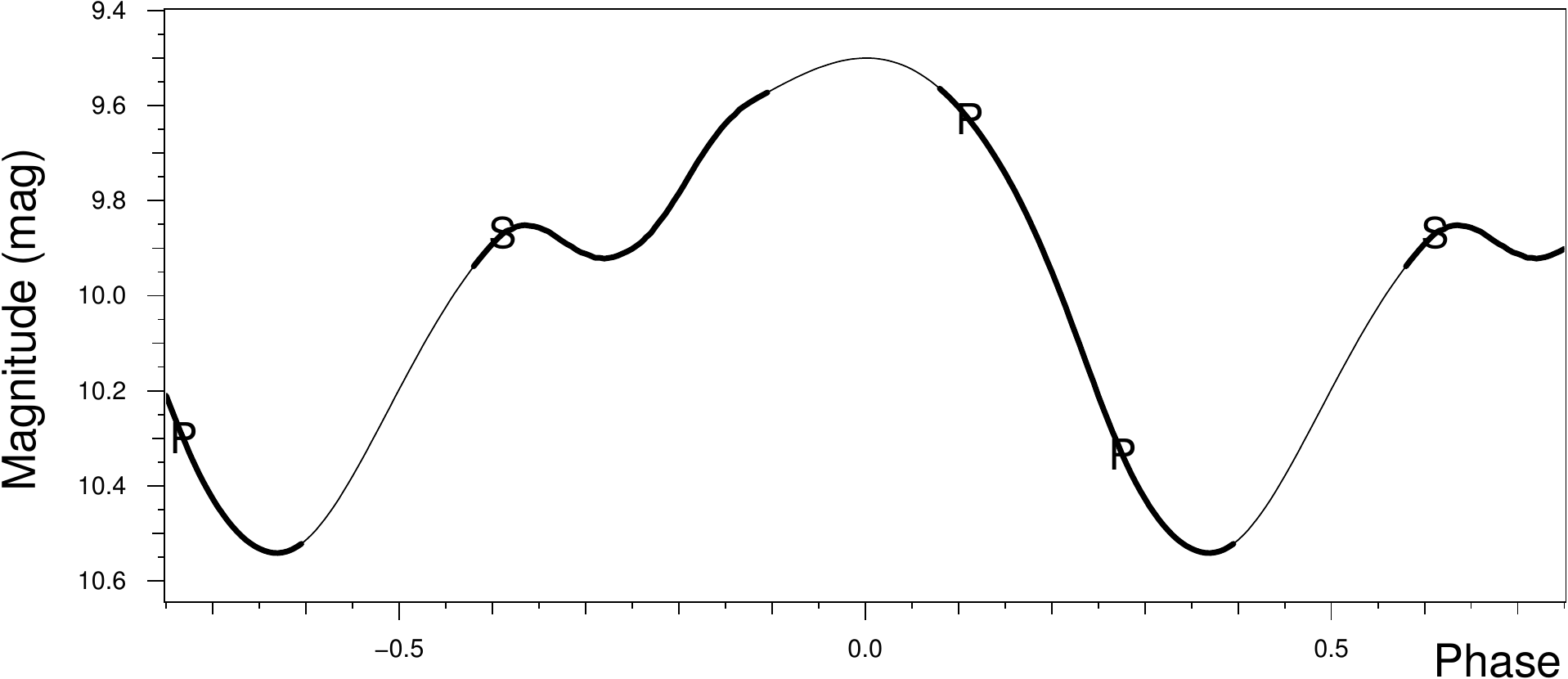} 
  \includegraphics[width=11cm]{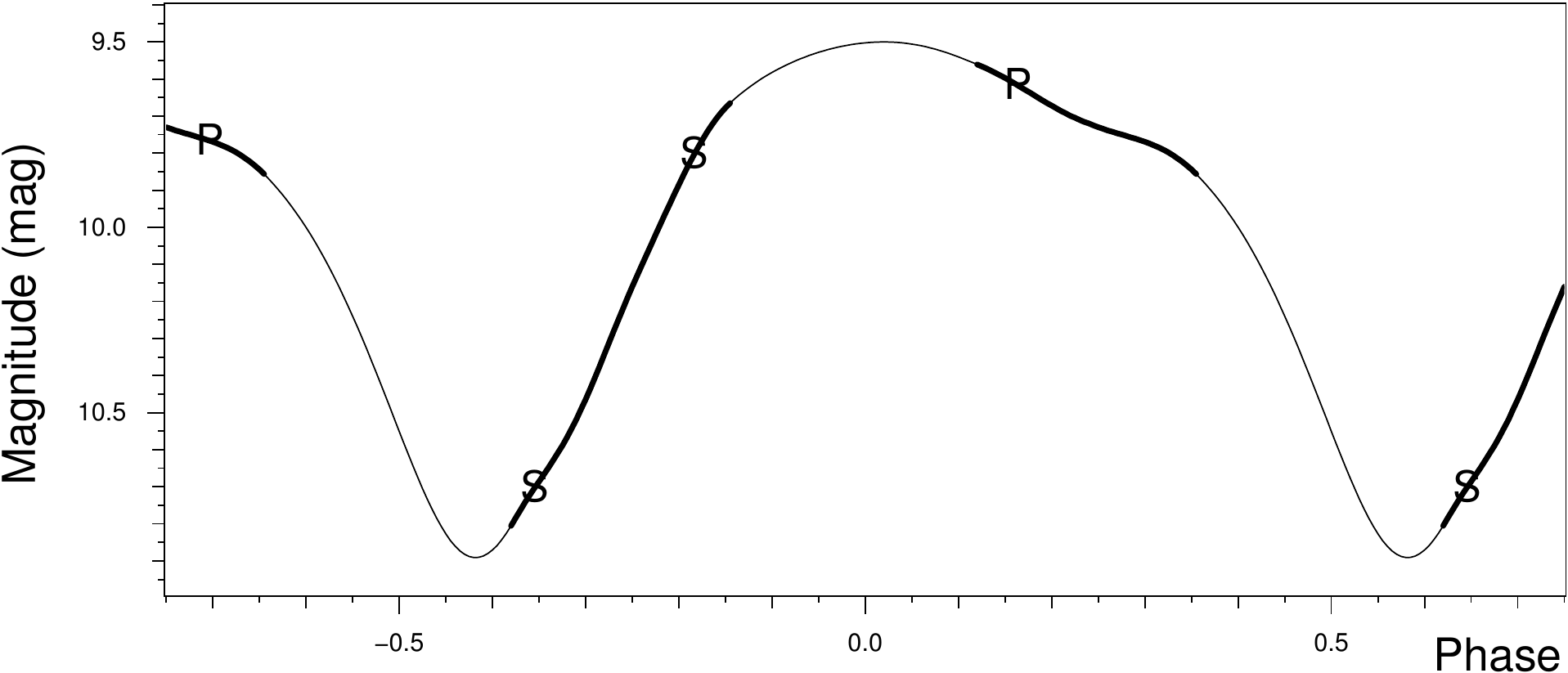}
  \caption{Examples of simulated light curves. The parts of the light curve where the primary (P) or the secondary (S) are eclipsed are labelled (P or S) and bold. The simulation parameters are given in table.~\ref{tabsimul}} 
  \label{Figsimul}
\end{figure}

\begin{deluxetable}{@{}l cccc}
\tablecaption{Parameters of the simulated light curves of Fig.~\ref{Figsimul} (from top to bottom)\label{tabsimul}}
\tablehead{
\colhead{Type:} & \colhead{Contact} & \colhead{Semidetached}& \colhead{Contact}& \colhead{Semidetached}}
\tablewidth{0pt}
\startdata
  Mass ratio & 0.1 & 0.1 & 0.1 & 0.1  \\
  Incidence & $ 65\degr$  & $ 65\degr$  & $ 65\degr$  & $ 65\degr$ \\
  Temperatures: \\
  Primary  & 2600 K & 2600 K & 2600 K & 2600 K\\
  Secondary  & 2800 K & 2600 K & 2800 K& 2600 K \\ 
  \hline
  Spot primary: \\
  Longitude & $90\degr$ & $90\degr$ & $60\degr$ & $120\degr$ \\
  Radius & $90\degr$ & $90\degr$ & $90\degr$ & $90\degr$  \\
  Dimfactor & $0.5$  & $0.5$& $0.5$ & $0.5$ \\
  \hline
  Spot secondary: & no & no & no & no\\ 
\enddata
\end{deluxetable}

\subsection{Models of eclipsing variables}

Since the pioneering work of \citet{wilson}, there are several codes to synthesise light curves of eclipsing variables for widely separated, semi--detached or contact binaries. We used the NightFall code (see above) and the PHOEBE\footnote{http://www.phoebe-project.org/} (Physics of Eclipsing Binaries) code. The first step is to establish the geometry of the system. Roche equipotentials are computed on the basis of the mass ratio of the two components. All that remains is to specify which components fills it Roche lobe. Knowledge of the period and the total mass of the system then allows to fix the system dimensions. The flux emerging from the system is obtained either by assuming that the two components radiate as black bodies or, in a more precise manner by using model atmosphere. The basic parameter is the actual effective temperature of each component. The effect of gravity brightening (the components do not have a spherical shape), the effect of limb darkening, the reflection of light from one component to the other must be taken into account. The light curve is due to the mutual eclipses of the components, their shapes, temperature differences, surface brightness (gravity brightening, limb darkening, reflection), and the presence of possible spots.

In a nutshell, there are 6 basic parameters : the mass ratio of the two components, their temperatures, their Roche lobe filling factor, the inclination of the orbit on the sky. Thus defined systems exhibit symmetric light curves: both maximum obtained when the two components appear further apart are the same. Both primary and secondary eclipses, although with different depths, exhibit symmetry about the eclipse centres. The light curves of Miras do not show these symmetries. It is therefore necessary to assume temperature inhomogeneities on the surface of the components. In the softwares we used, these differences are modelled by one or more spots of circular shape. One must specify on which component the spot is located (primary or secondary), the longitude and latitude of its centre, its radius and its temperature difference (positive or negative) from the rest of the star. 

Figure~\ref{FigRVCen} is the result of a fit assuming a spot on each component. Table~\ref{tabfitRV} gives the used parameters. It is worth noting that a fit on the visual light curve gives a fairly satisfactory result for the infrared light curves as well. 

Figure~\ref{Figsimul} presents four examples of simulated light curves achieved by varying the simulation parameters with a cool spot on the primary. For these examples three parameters vary: i) the system geometry (contact of semi-detached), ii) The temperature of the secondary, iii) the longitude of the spot. Table~\ref{tabsimul} gives the four set of parameters.

It is thus possible to reproduce light curves with a steep slope or with more or less pronounced bumps or with a concavity on the descending phase.

\section{Conclusions}

The consensus around the pulsation paradigm was clearly based on false assumptions, namely \emph{``Miras are aperiodic''}, probably because, at that time, few data were available and because periods were evaluated by the difference in dates between successive maximum rather than fitted on the whole data. Furthermore the dramatic shift of the pulsation mechanism from the centre of the star to the external layers was made without thoroughly reviewing the pulsation paradigm whereas the instability zone -- no longer at the centre but on the external layers -- makes the necessary synchronisation of the radial pulsation impossible. This mechanism change requires convection to be inhibited without explaining this inhibition. Recent 3D models show strong convective motions consistent with observations but no radial pulsation with large amplitudes as seen for the Miras.

Recent advances on red giants asteroseismology show rich pulsation spectra consistent with what theory predicts. By contrast, the variability of Miras with their single frequencies and their very high amplitudes can not be explained by a pulsation which would produce many simultaneous modes. This objection was already stressed by \citet{jeans} (see Appendix A) and is impossible to overcome.

Definitively rejecting the pulsation paradigm, we feel binarity offers the best prospects. The stability of the period is easily explained, as well as the shapes of the light curves and their variations. We found no fact that seriously challenges the presence of a companion.

A better understanding of these objects will lead to a better understanding of the interstellar medium as these objects, when in their last phase, eject the main part of their masses producing magnificent planetary nebulae contributing to the enrichment of heavy elements in the interstellar medium. They are also the brightest objects in the infrared. A better understanding will make it possible to use them as distance indicators provided we find a more precise relation between period and luminosity. It should be possible to apply these conclusions to the other objects thought to be radially pulsating, the RR Lyrae and the Cepheids in particular.

\acknowledgements
The author is indebted to the numerous observers whose patient observations have made this research possible. It is a pleasure to thank Eric Thi\'ebaut for his help and the provision of software written in Yorick language, Prof. Harm Habing for his encouragement and help and Prof. Alain Jorissen who carefully read this paper and suggest many improvements

\newpage
\appendix

\section{The historical development of the pulsation paradigm}

Cepheids were initially described as eclipsing variables. The study of their spectra cast doubt on this explanation. Such is the case of \citet{plummer} who estimated that the radial velocities deduced from the spectra of Cepheids were not compatible with a Keplerian motion and there was no evidence of a secondary spectrum. This leads the author to propose a motion of the atmosphere of the star from its centre. This view was shared by \citet{shapley}.

\subsection{First theory of Cepheids}

In 1918, Eddington proposes a first version of his 1-D theory of Cepheids \citep{eddington1918}. In this paper, he investigates the theory of a pulsating mass of gas. His theory is based on an adiabatic oscillation of the star leading to a period proportional to the inverse of the average density of the star. Applied to Delta Cephei it fixes a period as between 2.88 and 7.22 days. In this paper only radial pulsation is considered. \citet{eddington1918} is aware of this limit and justifies his choice because it is much simpler:
\emph{``The type of pulsation here considered is symmetrical about the centre; that is to say, the star remains spherical, but expands and contracts. It is possible that the actual oscillation may be an elliptical deformation; but I think that a symmetrical oscillation is more probable in a star of low density, and it is much simpler to investigate.''}.

\subsection{Jean's challenge}
Surprisingly, only \citet{jeans} emphasised this issue: \emph{``A more serious objection, however, is that a mass of gas has a large number of free vibrations whose periods are, in general, incommensurable. The free pulsations of a sphere of gas could only exhibit regularly recurring maxima in the very improbable event of all the vibrations except one having been damped to zero amplitude, while this particular one persisted with vigour.''}. 

\citet{jeans1926} is returning once again on this issue a year later without any success.

\subsection{The new theory of Cepheids}
In 1941 \citet{eddington1941} presented a last version of his theory of the Cepheids. Eddington thought the variation of luminosity resulted from a structure instability leading to periodic variations of radius and luminosity. According to \citet{eddington1941} some stars are unstable, and the instability mechanism is to be found at the star centre: a pressure rising strongly boosts nuclear reactions, generating more energy and leading to pressure and temperature increase and, consequently, inflating the star. This star inflating, in turn, leads to a fall in pressure and temperature at the centre, a slowing down of the nuclear reaction and consequently to a deflation of the star. This pattern being repeated leads to a periodic variation of star radius and light. From a simple evaluation of the free fall time, it is easy to show that period is proportional to the inverse of the square root of density (``$P \sqrt{\rho}$ \emph{law}'').

This paper has been a milestone in the debate on the origin of variation of Cepheids and Long-Period Variables. \citet{eddington1941} does not justify his assumption of a single mode of radial pulsation, as he did in 1918, neither he reported the well-founded Jean's objections.

\subsection{The Hoyle and Littleton challenge}
Two years after Eddington's paper, \citet{hoyle1943} proposed a new Cepheid Theory. They study a solution with a common atmosphere around the two components of a binary star. In fact, they echo Kuiper's idea explaining other variables namely $\beta$ Lyr -- this explanation is still accepted today -- as an interacting and eclipsing binary. In Hoyle and Lyttleton's model the common atmosphere shared by both stars orbiting one round the other doesn't take part in that common orbiting; it is thus subject to a kind of pulsation each time a star crosses the atmosphere. The authors note that following this model, they find again the $P \sqrt{\rho}$ \emph{law} previously found by Eddington. One of their main arguments is the absence of binaries among the known Cepheids if Eddington's theory is true.

The debate continued until 1949, led by \citet{code} who supported the pulsation paradigm.
 
For Code, the analysis applied to the Cepheids has also been successfully applied to the Miras and so the mechanism of long period variables (RV Tauri, Miras) is the same. But observations show that Miras and RV Tauri are not strictly periodic, this shared mechanism cannot be related to an orbital movement which must be strictly periodic.

Hoyle and Lyttleton fought until 1949, then the debate was given up and the pulsation paradigm for Cepheids, RR Lyrae and Mira variables was then widely accepted although deep changes occurred over the years. However no definitive arguments were put forward and neither thesis outweighed the other. Today some of these arguments are known to be wrong. In particular Miras are more strictly periodic than previously assumed.

\section{Light curves fitting results}
\footnotesize
\begin{longtable}{rcccccc}
\caption{\label{tabfit1}Light curves fitting results of objects with fixed periods or with a constant variation period rate. Dates are HJD-2450000.}
\\
\hline
\hline
 Star & Maximum & Minimum & Date maximum & Period & Asymmetry & Period variation\\
  & (mag) & (mag) & (HJD-2450000) & (days) & factor & (day/year)\\
 \hline
\endfirsthead
\hline
Star & Maximum & Minimum & Date maximum & Period & Asymmetry & Period variation\\
& (mag) & (mag) & (HJD-2450000) & (days) & factor & (day/year)\\
\hline
\endhead
\hline
\endfoot
\hline
\hline
\endlastfoot
  AL And &  11.1 &  14.4 &  51515.2 &   293.8 &  0.30 &  0.00\\
  RR And &   9.5 &  15.4 &  51603.2 &   327.9 &  0.50 &  0.00\\
   V And &   9.7 &  14.2 &  51540.1 &   258.4 &  0.47 &  0.00\\
   X And &   9.4 &  14.6 &  51703.7 &   345.6 &  0.36 &  0.00\\
   Y And &   9.5 &  14.7 &  51490.2 &   221.2 &  0.43 &  0.00\\
  YZ And &  10.8 &  15.1 &  51611.2 &   207.9 &  0.43 &  0.00\\
  AC Ant &  11.3 &  14.3 &  51555.9 &   215.8 &  0.47 &  0.00\\
   V Ant &   9.2 &  14.3 &  51548.9 &   302.2 &  0.28 &  0.00\\
   X Ant &   9.3 &  12.6 &  51507.4 &   162.2 &  0.46 &  0.00\\
   T Aps &   9.7 &  14.6 &  51564.0 &   261.4 &  0.39 &  0.00\\
   R Aql &   6.7 &  10.8 &  51551.5 &   273.8 &  0.53 & -0.48\\
  RT Aql &   9.1 &  14.2 &  51497.9 &   328.3 &  0.33 &  0.00\\
  RV Aql &   9.5 &  14.3 &  51597.0 &   218.7 &  0.43 &  0.00\\
  TU Aql &   9.7 &  14.9 &  51619.0 &   272.1 &  0.46 &  0.00\\
  TV Aql &  10.4 &  14.4 &  51557.9 &   241.8 &  0.53 &  0.00\\
   Z Aql &   9.6 &  13.1 &  51558.6 &   130.2 &  0.50 &  0.00\\
  RR Aqr &  10.0 &  13.5 &  51584.6 &   182.2 &  0.48 &  0.00\\
  RW Aqr &   9.8 &  13.7 &  51496.7 &   140.1 &  0.46 &  0.00\\
   S Aqr &   9.0 &  13.5 &  51506.2 &   279.7 &  0.37 &  0.00\\
   T Aqr &   7.8 &  12.5 &  51620.3 &   201.7 &  0.48 &  0.00\\
   U Ara &   9.0 &  13.6 &  51537.7 &   225.3 &  0.42 &  0.00\\
   T Ari &   8.5 &  10.2 &  51577.2 &   320.9 &  0.49 &  0.00\\
   U Ari &   8.8 &  14.6 &  51432.0 &   373.0 &  0.38 &  0.00\\
  AA Aur &  10.1 &  14.6 &  51463.9 &   268.5 &  0.49 &  0.06\\
  AZ Aur &  10.2 &  13.4 &  51598.1 &   414.8 &  0.44 &  0.00\\
  GO Aur &  10.3 &  14.9 &  51477.1 &   295.5 &  0.49 &  0.00\\
  GQ Aur &  11.2 &  15.3 &  51545.3 &   305.3 &  0.52 &  0.00\\
  RR Aur &  10.3 &  15.0 &  51424.1 &   310.3 &  0.46 &  0.00\\
  ST Aur &  11.4 &  14.6 &  51439.4 &   291.7 &  0.50 &  0.00\\
   V Aur &   9.7 &  12.3 &  51556.7 &   352.9 &  0.51 &  0.00\\
   R Boo &   7.3 &  12.2 &  51458.7 &   223.6 &  0.48 &  0.00\\
  RZ Boo &  10.5 &  11.7 &  51589.8 &   212.2 &  0.47 &  0.00\\
   U Boo &  10.6 &  11.8 &  51525.7 &   201.6 &  0.50 &  0.00\\
   R CMi &   8.2 &  10.6 &  51429.8 &   337.0 &  0.45 &  0.00\\
   R CVn &   7.9 &  11.6 &  51467.4 &   329.1 &  0.53 &  0.00\\
   V CVn &   7.3 &   7.8 &  51640.5 &   192.2 &  0.57 &  0.00\\
  RT Cam &  10.3 &  13.7 &  51490.4 &   365.2 &  0.59 &  0.00\\
  RY Cam &   7.8 &   8.5 &  51542.9 &   136.2 &  0.52 &  0.00\\
  SU Cam &  10.4 &  14.5 &  51681.7 &   285.1 &  0.43 &  0.00\\
  SW Cam &  10.4 &  14.1 &  51551.2 &   253.7 &  0.51 & -0.04\\
   W Cam &  10.9 &  14.5 &  51641.2 &   283.6 &  0.36 &  0.00\\
   X Cam &   8.1 &  12.4 &  51576.4 &   143.8 &  0.50 &  0.00\\
   R Car &   5.2 &   9.9 &  51530.6 &   308.3 &  0.48 &  0.00\\
   Z Car &  11.3 &  14.3 &  51367.8 &   388.0 &  0.43 &  0.00\\
  VZ Cas &  10.1 &  13.1 &  51624.0 &   169.3 &  0.47 &  0.00\\
  WY Cas &   9.2 &  15.1 &  51494.1 &   476.8 &  0.40 &  0.00\\
  RS Cen &   8.8 &  13.4 &  51495.2 &   164.8 &  0.42 &  0.00\\
  RT Cen &   9.4 &  12.2 &  51567.6 &   255.6 &  0.43 &  0.00\\
   U Cen &   8.6 &  12.7 &  51582.5 &   219.8 &  0.51 &  0.00\\
   W Cen &   8.9 &  13.3 &  51475.2 &   201.3 &  0.52 &  0.00\\
   X Cen &   8.4 &  13.5 &  51518.8 &   316.5 &  0.40 &  0.00\\
  AE Cep &  11.0 &  15.0 &  51522.0 &   168.7 &  0.49 &  0.00\\
  SZ Cep &   9.7 &  14.5 &  51501.7 &   327.3 &  0.51 &  0.00\\
   R Cet &   8.0 &  12.1 &  51569.0 &   166.0 &  0.41 &  0.00\\
   S Cet &   8.7 &  13.5 &  51569.3 &   319.6 &  0.55 &  0.00\\
   V Cnc &   8.0 &  12.5 &  51654.9 &   272.3 &  0.44 &  0.00\\
   V CrB &   8.4 &  11.3 &  51646.5 &   357.9 &  0.37 &  0.00\\
   W CrB &   8.8 &  12.9 &  51593.6 &   238.8 &  0.48 &  0.00\\
   X CrB &   9.3 &  13.5 &  51651.0 &   241.2 &  0.49 &  0.00\\
   R Crv &   8.1 &  13.2 &  51668.8 &   317.4 &  0.44 &  0.00\\
  BG Cyg &  10.1 &  11.9 &  51520.4 &   286.6 &  0.50 &  0.00\\
  BN Cyg &  11.4 &  15.0 &  51576.5 &   174.5 &  0.46 &  0.00\\
  BS Cyg &  10.6 &  14.1 &  51580.4 &   425.8 &  0.27 &  0.00\\
  BU Cyg &  10.7 &  14.5 &  51584.6 &   157.8 &  0.50 &  0.00\\
  CN Cyg &   9.3 &  14.1 &  51546.4 &   199.0 &  0.50 &  0.00\\
  CU Cyg &  10.7 &  13.9 &  51463.4 &   214.3 &  0.54 &  0.00\\
  DR Cyg &   9.7 &  14.9 &  51611.6 &   313.3 &  0.43 &  0.00\\
  FL Cyg &  11.5 &  14.0 &  51540.7 &   143.9 &  0.57 &  0.00\\
  RT Cyg &   7.3 &  11.5 &  51622.0 &   190.2 &  0.44 &  0.00\\
  TU Cyg &   9.7 &  14.0 &  51573.5 &   219.6 &  0.45 &  0.00\\
  WY Cyg &   9.7 &  14.6 &  51518.1 &   304.5 &  0.47 &  0.00\\
   Z Cyg &   8.8 &  13.2 &  51487.1 &   264.3 &  0.42 &  0.00\\
   S Del &   8.9 &  11.5 &  51585.8 &   279.5 &  0.47 &  0.00\\
   X Del &   9.6 &  15.1 &  51680.3 &   281.3 &  0.41 &  0.00\\
  SU Dor &   9.3 &  14.1 &  51587.0 &   235.7 &  0.43 &  0.00\\
   U Dor &   9.3 &  14.2 &  51583.7 &   393.5 &  0.23 &  0.00\\
   R Dra &   7.7 &  12.6 &  51423.0 &   246.0 &  0.44 &  0.00\\
  RT Dra &  10.2 &  13.8 &  51584.0 &   275.1 &  0.50 & -0.09\\
  RV Dra &  10.0 &  13.8 &  51463.1 &   208.7 &  0.37 &  0.00\\
   V Dra &  10.2 &  14.3 &  51426.8 &   278.6 &  0.52 &  0.00\\
   Y Dra &   9.5 &  14.9 &  51647.5 &   325.7 &  0.41 &  0.00\\
  RS Eri &   9.2 &  13.2 &  51422.0 &   301.3 &  0.46 &  0.00\\
   S Gem &   9.6 &  14.2 &  51480.8 &   292.5 &  0.43 &  0.00\\
   V Gem &   8.8 &  14.2 &  51506.3 &   275.3 &  0.45 &  0.00\\
  VX Gem &   8.9 &  12.9 &  51520.1 &   382.1 &  0.42 &  0.00\\
  AE Her &  10.1 &  14.7 &  51598.4 &   248.8 &  0.41 &  0.00\\
  AS Her &   9.1 &  13.7 &  51469.8 &   268.7 &  0.51 &  0.00\\
  AY Her &  11.0 &  12.8 &  51561.2 &   128.6 &  0.48 &  0.00\\
  DG Her &  10.8 &  14.5 &  51500.4 &   292.9 &  0.37 &  0.00\\
  RV Her &  10.5 &  15.0 &  51527.6 &   205.7 &  0.41 &  0.00\\
  RZ Her &  10.2 &  15.3 &  51399.9 &   328.7 &  0.45 &  0.00\\
  SY Her &   8.6 &  12.0 &  51558.7 &   116.9 &  0.51 &  0.00\\
   T Her &   8.0 &  12.5 &  51565.1 &   164.8 &  0.48 &  0.00\\
  TV Her &  10.1 &  14.7 &  51634.1 &   304.3 &  0.39 &  0.00\\
   W Her &   8.5 &  13.3 &  51475.9 &   279.5 &  0.47 &  0.00\\
  XZ Her &  10.9 &  14.2 &  51581.7 &   171.0 &  0.51 &  0.00\\
   X Hor &   8.7 &  10.3 &  51646.2 &   284.4 &  0.21 &  0.00\\
   R LMi &   7.9 &  12.9 &  51427.6 &   372.5 &  0.35 &  0.00\\
   S LMi &   8.6 &  13.9 &  51596.4 &   233.9 &  0.39 &  0.00\\
   U LMi &  11.4 &  12.2 &  51577.3 &   273.3 &  0.53 &  0.00\\
  SU Lac &  11.7 &  15.6 &  51543.3 &   308.1 &  0.47 &  0.00\\
   W Lac &  10.4 &  13.7 &  51561.9 &   322.4 &  0.49 &  0.00\\
  TZ Leo &  10.5 &  12.4 &  51436.9 &   327.3 &  0.60 &  0.00\\
  RR Lib &   9.0 &  14.2 &  51549.8 &   277.4 &  0.54 &  0.00\\
  RU Lib &   8.7 &  14.2 &  51673.0 &   315.9 &  0.49 &  0.00\\
   R Lup &  10.2 &  13.4 &  51604.1 &   234.4 &  0.50 &  0.00\\
   Y Lup &  10.5 &  14.8 &  51551.8 &   401.6 &  0.32 &  0.00\\
   R Lyn &   8.4 &  13.4 &  51610.3 &   378.6 &  0.46 &  0.00\\
  RU Lyn &  10.6 &  15.2 &  51535.0 &   243.7 &  0.41 &  0.00\\
   W Lyn &  10.4 &  15.1 &  51444.5 &   294.3 &  0.32 &  0.00\\
  RS Lyr &  10.8 &  14.9 &  51455.2 &   304.6 &  0.52 &  0.00\\
  RT Lyr &  10.8 &  15.0 &  51449.3 &   252.8 &  0.45 &  0.00\\
  RY Lyr &  10.2 &  14.9 &  51514.1 &   325.7 &  0.46 &  0.00\\
  RX Mic &  10.3 &  14.0 &  51544.1 &   237.8 &  0.48 &  0.00\\
   S Mic &   8.9 &  14.1 &  51535.6 &   208.8 &  0.39 &  0.00\\
  RS Mon &  10.1 &  14.8 &  51619.0 &   265.2 &  0.43 &  0.00\\
   T Nor &   7.9 &  13.1 &  51651.1 &   242.4 &  0.41 &  0.00\\
  RT Oct &  10.0 &  13.6 &  51499.5 &   178.8 &  0.44 &  0.00\\
  RU Oct &  11.6 &  13.7 &  51492.6 &   365.2 &  0.53 &  0.00\\
   T Oct &   9.3 &  14.1 &  51599.8 &   219.8 &  0.41 &  0.00\\
  AO Oph &  11.3 &  15.7 &  51489.8 &   218.1 &  0.49 &  0.00\\
   R Oph &   7.7 &  12.7 &  51579.8 &   303.2 &  0.42 &  0.00\\
  RR Oph &   9.4 &  14.0 &  51428.6 &   294.5 &  0.40 &  0.00\\
  RU Oph &   9.8 &  13.1 &  51553.3 &   201.7 &  0.48 &  0.00\\
  SS Oph &   9.0 &  13.5 &  51535.3 &   179.3 &  0.53 &  0.00\\
  RR Ori &  10.3 &  14.6 &  51564.7 &   252.7 &  0.46 &  0.00\\
   R Pav &   8.4 &  13.2 &  51475.8 &   230.3 &  0.47 &  0.00\\
  FF Peg &  10.6 &  14.5 &  51517.2 &   251.0 &  0.66 &  0.00\\
  RT Peg &  10.5 &  14.7 &  51603.8 &   215.2 &  0.46 &  0.00\\
  RW Peg &   9.9 &  13.7 &  51532.2 &   208.6 &  0.47 &  0.00\\
   S Peg &   8.4 &  12.8 &  51386.6 &   319.9 &  0.50 &  0.00\\
   V Peg &   9.1 &  14.8 &  51489.3 &   303.5 &  0.40 &  0.00\\
   Y Peg &  11.1 &  15.3 &  51526.9 &   206.7 &  0.46 &  0.00\\
  AS Pup &   8.3 &  11.1 &  51477.4 &   328.4 &  0.59 &  0.00\\
  RV Pup &   8.1 &  11.8 &  51544.3 &   188.1 &  0.54 &  0.00\\
  RW Pup &  10.2 &  13.7 &  51425.2 &   336.9 &  0.45 &  0.00\\
  TU Pup &  10.0 &  13.8 &  51545.4 &   241.1 &  0.53 &  0.00\\
   S Pyx &   9.4 &  14.2 &  51609.6 &   206.6 &  0.44 &  0.00\\
  RR Sco &   6.5 &  11.3 &  51551.2 &   279.2 &  0.51 &  0.00\\
  RS Sco &   7.4 &  12.1 &  51386.1 &   319.2 &  0.38 &  0.00\\
   X Sco &  11.0 &  14.2 &  51473.0 &   199.3 &  0.52 &  0.00\\
   R Ser &   6.9 &  13.2 &  51564.8 &   355.4 &  0.38 &  0.00\\
   R Sgr &   7.5 &  12.1 &  51416.1 &   268.7 &  0.48 &  0.00\\
  RR Sgr &   7.1 &  13.3 &  51467.2 &   334.2 &  0.41 &  0.00\\
  RT Sgr &   7.7 &  12.9 &  51609.0 &   306.8 &  0.49 &  0.00\\
  RU Sgr &   7.4 &  12.9 &  51516.1 &   240.5 &  0.40 &  0.00\\
  RW Sgr &   9.8 &  10.4 &  51542.1 &   187.9 &  0.48 &  0.00\\
  RX Sgr &  10.1 &  13.9 &  51458.4 &   333.2 &  0.46 &  0.00\\
   S Sgr &  10.7 &  13.8 &  51557.7 &   230.7 &  0.53 &  0.00\\
  RX Tau &  10.3 &  13.9 &  51616.6 &   334.4 &  0.44 &  0.00\\
  TZ Tau &  11.9 &  14.1 &  51455.0 &   268.2 &  0.45 &  0.00\\
  NT Tel &  11.2 &  14.3 &  51634.9 &   254.9 &  0.55 &  0.00\\
   R Tri &   6.5 &  11.4 &  51628.3 &   267.1 &  0.46 &  0.00\\
   R UMa &   7.7 &  12.8 &  51610.1 &   301.6 &  0.34 &  0.00\\
  RR UMa &   9.8 &  14.5 &  51489.2 &   231.0 &  0.43 &  0.00\\
  RS UMa &   9.5 &  14.2 &  51467.0 &   259.5 &  0.45 &  0.00\\
  RU UMa &   8.9 &  14.5 &  51441.7 &   250.9 &  0.33 &  0.00\\
   T UMa &   7.8 &  12.8 &  51515.5 &   256.4 &  0.37 &  0.00\\
  RS Vir &   8.3 &  13.7 &  51415.0 &   354.2 &  0.29 &  0.00\\
   U Vir &   8.3 &  12.1 &  51580.2 &   206.9 &  0.50 &  0.00\\
   V Vir &   9.2 &  13.8 &  51636.1 &   249.6 &  0.35 &  0.00\\
   Y Vir &   9.7 &  13.8 &  51459.7 &   217.8 &  0.48 &  0.00\\
   R Vul &   8.2 &  12.2 &  51601.6 &   136.7 &  0.49 &  0.00\\
\end{longtable}

\begin{landscape}
\scriptsize
\begin{longtable}{rcccccccccccccc}
\caption{\label{tabfit2}Light curves fitting results of objects with variable periods. Dates are HJD-2450000.}
\\
\hline
\hline
 Star & Max. & Min. & Date max. & Period & Date & Period & Date & Period & Date & Period & Date & Period & Date & Period \\
      & (mag) & (mag) & HJD-2450000 & (days) & &  &  &  & &  &  &  & &  \\
\hline
\endfirsthead
\hline
 Star & Max. & Min. & Date max. & Period & Date & Period & Date & Period & Date & Period & Date & Period & Date & Period \\
      & (mag) & (mag) & HJD-2450000 & (days) & &  &  &  & &  &  &  & &  \\
\hline
\endhead
\hline
\endfoot
\hline
\hline
\endlastfoot
   R And &   7.6 &  14.4 &  51363.5 &   411.7 &  46842.2 &   406.5 &  43875.6 &   419.8 &  35389.7 &   400.5 &  31184.1 &   414.4 &  20467.4 &   406.1  \\
   T And &   8.9 &  13.9 &  51398.3 &   278.0 &  40789.7 &   286.1 &  35969.7 &   276.4 &  31575.9 &   280.6  \\
  RS Aql &  11.0 &  14.3 &  51454.8 &   396.7 &  54415.8 &   434.4 &  38163.6 &   420.3  \\
   X Aql &   9.5 &  15.7 &  51469.8 &   343.9 &  48012.2 &   336.7 &  45248.6 &   351.1 &  40685.6 &   346.1  \\
  RS Aqr &  10.6 &  13.4 &  51582.4 &   217.7 &  49885.1 &   227.7 &  44159.1 &   210.7 &  41874.7 &   217.7  \\
   R Aur &   8.2 &  13.1 &  51343.5 &   457.0 &  30234.5 &   453.9 &  29216.2 &   466.2  \\
  RU Aur &  10.3 &  15.5 &  51547.2 &   463.8 &  45904.1 &   483.1 &  38973.9 &   460.0 &  33672.2 &   469.5  \\
  VX Aur &   8.7 &  12.7 &  51546.9 &   330.5 &  52574.8 &   322.4 &  42907.3 &   329.4  \\
   X Aur &   8.6 &  12.2 &  51547.7 &   166.6 &  39230.1 &   162.1 &  28564.9 &   166.3 &  23067.9 &   162.6  \\
   S Boo &   8.6 &  13.2 &  51516.2 &   274.4 &  29119.7 &   268.6 &  22760.6 &   275.8  \\
   T CMi &  10.4 &  14.3 &  51522.3 &   316.9 &  46066.0 &   313.6 &  35659.1 &   334.2 &  27737.9 &   307.2 &  16366.6 &   334.3  \\
   U CMi &   8.8 &  12.1 &  51392.4 &   414.6 &  46615.6 &   403.7 &  44600.3 &   419.5 &  40498.8 &   429.5 &  34341.0 &   391.6 &  31016.7 &   420.2 \\
   R Cam &   8.4 &  12.4 &  51527.3 &   270.4 &  31394.5 &   270.8 &  28188.9 &   258.1 &  25564.8 &   273.9  \\
   T Cam &   8.1 &  12.8 &  51642.8 &   376.4 &  39634.5 &   366.0 &  36315.4 &   380.1 &  24150.0 &   374.2  \\
  RW Car &   9.7 &  14.6 &  51367.4 &   316.9 &  39078.4 &   317.7 &  31421.8 &   324.8 &  22755.1 &   311.2  \\
  RY Car &  12.1 &  14.1 &  51330.7 &   428.5 &  46596.3 &   407.5 &  41766.8 &   429.0 &  33004.5 &   418.1  \\
   R Cas &   6.7 &  12.2 &  51375.0 &   430.2 &  44601.8 &   433.6 &  39741.2 &   422.9 &  33766.1 &   435.2 &  20019.4 &   425.8  \\
  SS Cas &  10.2 &  13.2 &  51565.5 &   139.5 &  47005.3 &   138.1 &  42661.5 &   140.2  \\
   T Cas &   8.0 &  11.3 &  51648.4 &   436.5 &  37448.8 &   451.7 &  29550.2 &   434.1 &  28472.1 &   446.7  \\
   V Cas &   8.0 &  12.2 &  51676.5 &   231.8 &  48404.1 &   225.8 &  40217.0 &   233.3 &  29158.8 &   225.2 &  23711.9 &   227.2  \\
   X Cas &  10.3 &  12.0 &  51605.4 &   419.7 &  45536.0 &   431.0 &  27793.8 &   414.4 &  22745.9 &   439.0  \\
   Y Cas &  10.4 &  14.9 &  51540.0 &   421.0 &  43504.9 &   404.1 &  39016.7 &   417.2 &  26623.7 &   407.7 &  20587.1 &   426.8  \\
   Z Cas &  10.9 &  15.4 &  51405.0 &   490.9 &  50591.0 &   505.9 &  46156.9 &   507.3 &  42825.2 &   488.5 &  28892.3 &   504.7 &  25508.1 &   491.0  \\
   R Cen &   6.5 &   9.2 &  51543.8 &   512.2 &  56350.0 &   490.0 &  36769.4 &   554.3 &  25297.9 &   539.0 &  21576.4 &   562.4  \\
   S Cep &   8.4 &  10.4 &  51566.1 &   479.1 &  43462.2 &   490.7 &  36584.0 &   478.7 &  31262.5 &   499.8 &  23457.2 &   474.3 &  19314.9 &   491.0  \\
   T Cep &   6.3 &  10.1 &  51740.5 &   389.6 &  48643.3 &   404.5 &  35691.7 &   375.1 &  29236.9 &   400.0 &  18705.4 &   386.0  \\
   W Cet &   8.3 &  13.3 &  51507.1 &   350.8 &  50463.0 &   341.6 &  35567.1 &   359.2 &  26861.8 &   346.2  \\
   Z Cet &   9.3 &  13.3 &  51473.1 &   184.2 &  55071.3 &   182.9 &  31951.6 &   185.4  \\
 omi Cet &   3.8 &   9.6 &  51489.3 &   332.0 &  43169.8 &   334.8 &  16985.1 &   330.5  \\
   R Cnc &   7.2 &  11.3 &  51472.4 &   362.9 &  31720.3 &   359.9 &  29018.5 &   366.4 &  24150.8 &   372.6  \\
  BH Cru &   7.3 &   9.8 &  51730.9 &   524.1 &  50377.0 &   538.6 &  40539.9 &   417.2  \\
  LX Cyg &  10.3 &  13.3 &  51543.9 &   585.2 &  56344.4 &   593.0 &  44212.6 &   477.9  \\
   R Cyg &   8.2 &  13.9 &  51405.7 &   420.2 &  44922.9 &   429.3 &  22481.5 &   427.3  \\
  RS Cyg &   7.4 &   8.6 &  50680.5 &   425.9 &  36977.6 &   411.4 &  29763.6 &   424.0 &  20156.7 &   413.7  \\
  RZ Cyg &  11.3 &  12.8 &  51383.7 &   543.7 &  50184.9 &   517.8 &  42618.4 &   554.0 &  29510.1 &   519.2 &  21833.9 &   568.4  \\
  ST Cyg &  10.2 &  13.8 &  51498.2 &   333.7 &  29439.9 &   342.4 &  22878.0 &   329.9  \\
  TY Cyg &  10.1 &  14.2 &  51441.0 &   359.2 &  37021.8 &   342.3 &  31829.5 &   360.9 &  25328.3 &   344.3  \\
   U Cyg &   7.6 &  10.5 &  51515.6 &   465.8 &  40222.4 &   458.8 &  31598.5 &   479.9 &  23440.1 &   449.1 &  19100.6 &   468.4  \\
 chi Cyg &   5.7 &  13.7 &  51531.3 &   404.2 &  40835.2 &   413.5 &  37008.7 &   403.6 &  26479.7 &   412.7 &  18997.8 &   403.2  \\
  ES Del &  11.0 &  15.1 &  51838.5 &   478.8 &  51838.4 &   511.6 &  38256.6 &   495.6  \\
  RX Del &  10.9 &  14.7 &  42305.8 &   183.1 &  47685.6 &   186.0  \\
   T Del &  10.0 &  14.9 &  51569.6 &   326.7 &  46874.3 &   332.1  \\
   U Dra &  10.2 &  13.9 &  51429.6 &   320.7 &  48087.8 &   329.2 &  41988.2 &   313.2 &  35097.9 &   311.9 &  28384.8 &   324.8 &  22595.1 &   313.5  \\
  WZ Dra &   9.8 &  13.9 &  51718.9 &   410.0 &  55914.9 &   411.3  \\
   ST Gem &   9.6 &  14.4 &  51435.3 &   247.3 &  47182.7 &   240.8 &  44533.4 &   247.4  \\
   T Gem &   8.5 &  13.3 &  51577.6 &   288.7 &  43102.9 &   284.8 &  38884.8 &   291.4 &  29396.6 &   282.6 &  25465.8 &   294.4 &  20331.1 &   284.2  \\
   T Gru &   8.7 &  11.1 &  51525.9 &   136.5 &  42888.2 &   137.3 &  33895.9 &   136.4  \\
   R Her &   9.3 &  15.3 &  51410.7 &   324.4 &  49294.4 &   315.4 &  39610.1 &   317.4  \\
  RS Her &   8.0 &  12.2 &  51524.6 &   217.6 &  43051.5 &   220.0  \\
  RU Her &   8.8 &  13.8 &  51800.6 &   485.6 &  50357.0 &   501.8 &  42059.4 &   473.9 &  38941.3 &   494.8 &  31831.1 &   479.5 &  25109.2 &   497.2 \\
   S Her &   7.7 &  12.6 &  51504.3 &   307.1 &  37036.6 &   306.2 &  27477.9 &   316.5 &  21088.4 &   301.2  \\
  SS Her &   9.1 &  12.4 &  51531.4 &   106.4 &  46376.0 &   107.7 &  43517.1 &   105.6 &  29438.2 &   108.2 &  26637.4 &   105.4 &  21015.1 &   109.3  \\
   R Hor &   6.4 &  13.4 &  51575.5 &   403.3 &  47247.4 &   413.2 &  44272.0 &   388.4 &  40668.5 &   410.4 &  24538.1 &   401.4 &  19482.3 &   409.1  \\
   T Hor &   8.5 &  13.0 &  51519.0 &   219.7 &  41586.9 &   215.1 &  36522.9 &   220.5 &  29020.0 &   213.7 &  25981.2 &   218.4  \\
   R Hya &   5.3 &   8.8 &  51707.7 &   385.7 &  27755.6 &   391.3 &  27479.2 &   419.1 &  21903.1 &   406.4  \\
  RR Hya &  10.3 &  13.5 &  51445.9 &   342.8 &  36356.3 &   333.8  \\
   S Hya &   8.0 &  12.3 &  51518.3 &   254.2 &  46537.2 &   260.9 &  37215.1 &   254.8 &  24829.3 &   259.8  \\
   T Hya &   8.0 &  12.0 &  51454.2 &   290.3 &  47122.2 &   277.8 &  40931.4 &   304.6 &  36689.1 &   288.2 &  27284.8 &   282.3 &  25174.7 &   291.5  \\
   W Hya &   6.5 &   9.3 &  51768.5 &   390.2 &  50679.2 &   367.6 &  44657.2 &   392.7 &  43357.5 &   376.5 &  34121.6 &   408.9  \\
   R Lac &   9.7 &  15.0 &  51492.5 &   298.1 &  44968.2 &   306.6 &  40793.2 &   295.7 &  39581.4 &   300.3  \\
   S Lac &   8.5 &  13.0 &  51485.7 &   242.0 &  49371.0 &   238.7 &  41976.1 &   241.8 &  30510.9 &   236.1 &  25957.9 &   243.4 &  19866.2 &   240.8  \\
   R Leo &   6.2 &   9.9 &  51676.2 &   309.4 &  45581.3 &   315.3 &  40443.2 &   308.5 &  33367.0 &   313.7 &  28676.2 &   308.1 &  20183.3 &   317.9  \\
   W Leo &  10.6 &  14.7 &  51405.9 &   375.5 &  49114.6 &   393.2 &  30307.9 &   385.0  \\
   R Lep &   7.6 &   9.7 &  51702.8 &   440.2 &  36224.2 &   427.6 &  31545.6 &   448.0 &  19205.2 &   426.5  \\
   T Lep &   8.7 &  12.8 &  51470.7 &   374.3 &  44274.5 &   362.0 &  41153.4 &   376.8 &  38226.9 &   364.3 &  29677.8 &   372.7 &  23488.0 &   365.9  \\
  RS Lib &   7.9 &  12.0 &  51534.4 &   218.5 &  48421.9 &   220.1 &  45856.1 &   215.7 &  38331.5 &   221.3 &  32202.9 &   215.0 &  23031.5 &   219.6  \\
   U Lyn &  10.6 &  14.7 &  51606.4 &   439.8 &  45520.6 &   426.3 &  39719.4 &   443.8 &  29044.5 &   428.6 &  20935.1 &   443.8  \\
  SS Lyr &  10.6 &  14.3 &  51464.4 &   348.8 &  47716.5 &   360.1 &  43277.3 &   343.6  \\
   U Lyr &  10.0 &  11.8 &  51646.9 &   448.1 &  43985.5 &   460.5 &  33718.2 &   447.8 &  26952.5 &   462.0  \\
   W Lyr &   7.9 &  11.8 &  51595.2 &   197.2 &  48994.9 &   195.6 &  45004.7 &   199.2 &  33103.0 &   194.7 &  29732.0 &   199.2 &  23953.1 &   194.4  \\
   Z Lyr &  10.6 &  14.9 &  51280.5 &   287.6 &  48432.7 &   282.3 &  43472.4 &   295.1 &  32006.0 &   281.8 &  26653.0 &   294.1 &  18860.8 &   283.2  \\
   R Mic &   9.5 &  13.4 &  51263.5 &   138.6 &  45507.0 &   140.3 &  42688.7 &   137.9 &  34566.0 &   141.1 &  34564.2 &   137.1 &  20611.6 &   139.5  \\
   V Mon &   7.5 &  13.2 &  51307.1 &   333.5 &  48460.4 &   330.8 &  43279.7 &   341.5 &  37221.8 &   328.6 &  30693.6 &   338.9 &  27815.3 &   333.6  \\
   Y Mon &   9.4 &  14.0 &  51441.8 &   227.2 &  48069.9 &   231.4 &  37415.8 &   227.2 &  27937.5 &   234.8 &  21813.7 &   227.7  \\
   R Nor &   8.0 &  12.0 &  49594.0 &   518.1 &  51789.8 &   496.9 &  43254.9 &   475.8 &  39530.8 &   513.2 &  35599.5 &   495.6  \\
   U Oct &   8.6 &  13.6 &  51271.9 &   301.2 &  42284.4 &   303.8 &  23528.4 &   302.2  \\
   X Oph &   7.2 &   8.8 &  51442.8 &   341.7 &  43279.8 &   326.3 &  35403.5 &   336.7 &  28071.5 &   328.6 &  22797.2 &   338.2  \\
   Z Oph &   8.2 &  12.2 &  51699.2 &   348.3 &  44257.6 &   352.2 &  38584.8 &   348.3 &  28452.3 &   347.2 &  20929.9 &   350.2  \\
   R Ori &   9.8 &  13.0 &  51330.2 &   378.1 &  40003.0 &   385.6 &  37238.7 &   374.0 &  21378.1 &   381.1  \\
   S Ori &   8.8 &  12.6 &  52783.2 &   432.9 &  43955.1 &   406.8 &  36498.8 &   446.0 &  26796.0 &   400.8 &  21031.6 &   420.7  \\
   U Ori &   7.2 &  12.2 &  51504.0 &   377.2 &  48775.5 &   359.5 &  44835.5 &   374.1 &  27028.8 &   369.6 &  26300.9 &   380.2 &  18429.5 &   368.6  \\
   S Pav &   7.5 &   8.7 &  51622.7 &   390.0 &  18110.3 &   376.3  \\
   W Pav &   9.6 &  14.3 &  51391.6 &   278.9 &  40014.5 &   284.3 &  23284.5 &   282.9  \\
   R Peg &   8.2 &  13.0 &  51456.2 &   380.0 &  42098.5 &   369.4 &  40973.4 &   383.2 &  30033.8 &   375.1 &  25807.6 &   382.3 &  17717.3 &   376.0  \\
  RS Peg &  10.4 &  14.3 &  51453.1 &   420.1 &  34318.6 &   410.1 &  29903.4 &   415.7 &  20968.8 &   407.3  \\
  SS Peg &   9.5 &  14.0 &  45205.9 &   431.9 &  54219.9 &   400.0 &  40335.7 &   414.4  \\
  SX Peg &   8.9 &  12.8 &  51668.4 &   306.0 &  48776.2 &   312.9 &  44253.0 &   301.9  \\
   T Peg &   9.7 &  14.7 &  51549.8 &   368.5 &  45778.1 &   381.3 &  37025.8 &   367.6 &  29229.4 &   380.7 &  24449.7 &   359.3  \\
   W Peg &   8.7 &  12.3 &  51400.7 &   339.5 &  45610.9 &   345.6 &  30654.8 &   350.0 &  27888.2 &   339.0 &  20678.6 &   345.6  \\
   Z Peg &   8.7 &  13.2 &  51682.7 &   324.9 &  43285.3 &   336.4 &  40215.8 &   323.0 &  21354.0 &   330.1  \\
  RZ Per &   9.6 &  13.3 &  51532.8 &   349.8 &  46872.0 &   357.3 &  22932.5 &   352.9  \\
   U Per &   8.0 &  10.8 &  51424.2 &   317.0 &  39224.4 &   321.5 &  36283.8 &   316.6 &  31262.4 &   330.8 &  19996.6 &   316.6  \\
   Y Per &   8.9 &  10.0 &  51397.9 &   244.2 &  37349.8 &   253.5  \\
   W Pup &   8.3 &  12.1 &  51563.5 &   120.2 &  43357.9 &   119.8 &  37947.2 &   120.6  \\
   R Scl &   6.5 &   8.1 &  51773.5 &   372.6 &  29252.2 &   374.1 &  29247.1 &   389.2  \\
   S Scl &   7.2 &  12.9 &  51538.2 &   370.1 &  36513.2 &   360.7 &  28234.8 &   373.3 &  23987.0 &   354.2 &  20371.6 &   368.3  \\
   T Scl &   9.1 &  12.4 &  51413.0 &   206.8 &  43066.5 &   203.5 &  20167.5 &   200.2  \\
   U Scl &  10.1 &  15.2 &  51465.7 &   330.3 &  50430.7 &   338.4 &  44044.3 &   330.5 &  41316.1 &   343.2 &  40210.7 &   329.4 &  28129.1 &   336.2 \\
  RU Sco &   9.3 &  12.6 &  51612.0 &   352.1 &  39316.1 &   374.1 &  19756.5 &   362.8  \\
  RZ Sco &   8.9 &  11.2 &  51485.7 &   164.6 &  49036.6 &   161.7 &  39268.3 &   155.0 &  37733.1 &   164.3 &  32719.1 &   155.8 &  31245.5 &   165.5 \\
  SV Sco &  10.1 &  13.8 &  51392.2 &   258.6 &  53205.8 &   255.2 &  47438.8 &   251.3 &  42667.7 &   260.1 &  25153.6 &   253.5  \\
   Z Sco &   9.8 &  12.2 &  51585.4 &   355.9 &  41785.4 &   346.2 &  40716.9 &   341.9 &  31448.4 &   344.5 &  21957.3 &   357.0  \\
   S Sex &   9.3 &  12.9 &  51284.8 &   254.1 &  56300.0 &   258.6 &  47720.3 &   253.3 &  44868.0 &   269.4 &  35908.3 &   251.2 &  33044.6 &   263.4  \\
   T Sgr &   8.7 &  12.7 &  51482.0 &   392.4 &  46765.1 &   384.1 &  39713.5 &   399.3 &  33730.3 &   393.4  \\
  RU Tau &  11.1 &  14.6 &  51871.4 &   602.9 &  56300.0 &   604.7 &  41622.6 &   535.5 &  30983.2 &   614.0 &  23691.7 &   536.8  \\
   S Tau &  10.9 &  14.7 &  51655.5 &   367.9 &  43128.9 &   382.4 &  39522.8 &   367.2 &  36474.0 &   377.0 &  27515.4 &   374.7  \\
   Z UMa &   7.3 &   8.4 &  54384.0 &   187.7 &  39040.7 &   197.8 &  31028.2 &   181.0 &  22286.0 &   211.8  \\
   S UMi &   8.4 &  11.9 &  51451.3 &   320.8 &  43832.6 &   331.6 &  34155.1 &   322.0 &  28169.0 &   336.7 &  20916.0 &   320.1  \\
   T UMi &  10.2 &  12.8 &  51476.4 &   248.8 &  56300.0 &   197.2 &  42705.8 &   319.2 &  32399.2 &   307.7 &  27339.0 &   315.9  \\
   U UMi &   8.5 &  11.6 &  51562.9 &   319.2 &  41363.3 &   333.6 &  37601.2 &   318.7 &  35917.7 &   331.8 &  32731.8 &   316.1 &  28476.7 &   330.5  \\
   R Vir &   7.0 &  10.8 &  51551.9 &   146.6 &  48282.5 &   145.2 &  39649.7 &   147.3 &  36286.6 &   143.7 &  32117.6 &   149.1 &  29120.4 &   145.4  \\
  RU Vir &  10.3 &  12.5 &  51254.2 &   442.6 &  36420.0 &   427.5 &  32774.4 &   454.8 &  30773.0 &   430.6 &  24554.1 &   437.7  \\
   S Vir &   7.4 &  12.6 &  51489.8 &   375.3 &  45277.8 &   381.1 &  37800.3 &   371.2 &  32439.6 &   381.0  \\
  SU Vir &  10.2 &  14.1 &  51532.9 &   206.4 &  39316.5 &   209.3 &  25339.9 &   210.5  \\
   S Vol &   9.7 &  14.2 &  51425.1 &   389.2 &  41573.5 &   400.1 &  36564.7 &   381.4 &  33344.4 &   406.1 &  24505.4 &   390.8  \\
  RU Vul &   9.2 &   9.8 &  34650.4 &   155.4 &  52349.7 &   110.4  \\
\end{longtable}
\end{landscape}

\begin{longtable}{rcccccc}
\caption{\label{tabfit3}Light curves fitting results of SRb type variables with fixed periods or with a constant variation period rate. Dates are HJD-2450000.}
\\
\hline
\hline
Star & Maximum & Minimum & Date maximum & Period & Asymmetry & Period variation\\
& (mag) & (mag) & (HJD-2450000) & (days) & factor & (day/year)\\
\hline
\endfirsthead
 \hline
Star & Maximum & Minimum & Date maximum & Period & Asymmetry & Period variation\\
& (mag) & (mag) & (HJD-2450000) & (days) & factor & (day/year)\\
\hline
\endhead
\hline
\endfoot
\hline
\hline
\endlastfoot
  TY And &   9.6 &   9.8 &  51458.9 &   258.5 &  0.77 &  0.00\\
  UX And &   8.6 &   9.1 &  51509.3 &   219.5 &  0.57 &  0.00\\
   V Aqr &   8.5 &   8.9 &  51508.0 &   241.1 &  0.50 & -0.00\\
   S Aql &   9.4 &  11.0 &  51489.1 &   146.6 &  0.52 &  0.00\\
  RS Cam &   8.9 &   8.6 &  51504.7 &    88.6 &  0.54 &  0.00\\
  WZ Cas &   7.2 &   7.5 &  51567.4 &   371.0 &  0.28 & -0.03\\
   T Cnc &   8.7 &   9.4 &  51464.4 &   486.9 &  0.55 &  0.00\\
  TT CrB &  11.0 &  11.5 &  51562.4 &    69.3 &  0.42 &  0.00\\
   W Cyg &   5.8 &   6.2 &  51566.2 &   132.2 &  0.52 &  0.00\\
  AB Cyg &   7.6 &   8.0 &  51615.5 &   528.5 &  0.70 &  0.00\\
  AF Cyg &   7.1 &   7.3 &  51567.3 &    94.0 &  0.51 &  0.00\\
  EU Del &   6.2 &   6.2 &  51540.7 &    59.8 &  0.58 &  0.00\\
   R Dor &   5.6 &   5.9 &  51641.9 &   339.1 &  0.59 &  0.00\\
  AH Dra &   7.6 &   8.1 &  51607.8 &   190.0 &  0.38 &  0.00\\
  SY For &  10.3 &  11.3 &  51476.5 &   149.1 &  0.62 &  0.00\\
  RR Her &   8.5 &   9.1 &  51532.2 &   238.1 &  0.50 &  0.00\\
  ST Her &   7.3 &   7.8 &  51603.4 &   151.2 &  0.27 &  0.00\\
  RT Hya &   7.7 &   8.6 &  51504.0 &   246.3 &  0.29 & -0.11\\
   R Men &   8.2 &   9.7 &  51503.6 &   240.4 &  0.39 &  0.00\\
   T Mic &   8.6 &   6.9 &  51466.3 &   352.0 &  0.55 &  0.00\\
  FX Ori &   9.2 &   9.9 &  51353.9 &   690.6 &  0.71 &  0.00\\
  UV Pav &  10.9 &  12.6 &  51569.4 &   182.1 &  0.39 &  0.00\\
  DY Per &  11.0 &  12.9 &  51309.0 &   793.0 &  0.57 &  0.00\\
  RZ Sgr &   7.4 &  10.6 &  51622.1 &   208.2 &  0.38 &  0.00\\
  TV Sco &  10.8 &  10.9 &  51516.2 &   198.8 &  0.67 &  0.00\\
  RX UMa &  10.8 &  10.6 &  51603.2 &   195.2 &  0.97 &  0.00\\
  RY UMa &   7.5 &   7.2 &  51439.5 &   307.5 &  0.46 &  0.00\\
   R UMi &   9.2 &  10.1 &  51510.9 &   324.4 &  0.49 &  0.00\\
  $\mathrm{L}_2$ Pup &   4.9 &   5.5 &  51541.6 &   137.6 &  0.50 &  0.00\\
 $\pi_1$ Gru &   6.2 &   7.0 &  51508.8 &   194.8 &  0.48 &  0.00\\
\end{longtable}
\newpage

\end{document}